\pdfoutput=1

\documentclass{svjour3}                     
\smartqed  
\usepackage{graphicx}
\usepackage{natbib}
\usepackage{times}
\usepackage{comment}
\usepackage{fancyvrb}
\usepackage{boxedminipage}
\usepackage{url}
\usepackage{graphicx}
\usepackage{lscape}
\usepackage{latexsym}
\usepackage{listings}
\usepackage{rotating}
\usepackage{amsmath}
\lstdefinelanguage{xbgf}{%
  numbers=none,
  classoffset=0,
  morekeywords={a,g,n,l,p,t,s,x,v,*,+,?,true,fail},
  classoffset=1,
  morekeywords={%
		unfold,fold,inline,extract,abridge,detour,unchain,chain,%
		massage,distribute,factor,deyaccify,yaccify,eliminate,introduce,import,vertical,horizontal,rassoc,lassoc,%
		add,appear,widen,upgrade,unite,%
		remove,disappear,narrow,downgrade,%
		abstractize,concretize,permute,%
		define,undefine,redefine,inject,project,replace,%
		designate,unlabel,deanonymize,anonymize,%
		renameL,renameN,renameS,renameT,reroot,%
		dump},keywordstyle=\itshape\bfseries,
  classoffset=0,
  morestring=[b]',
  columns=flexible,
  basicstyle=\mdseries,
}

\lstdefinelanguage{bgf}{%
  numbers=none,
  morekeywords={a,g,n,l,p,t,s,x,v,*,+,?,string,int,true,fail},
  morestring=[b]',
  columns=flexible,
}

\lstdefinelanguage{smallbgf}{%
  numbers=none,
  morekeywords={a,g,n,l,p,t,s,x,v,*,+,?,string,int,true,fail},
  morestring=[b]',
  columns=flexible,
  basicstyle=\scriptsize\mdseries,
}

\lstdefinelanguage{fl}{%
  numbers=none,
  morekeywords={if,then,else,==,=,+,-},
}

\lstdefinelanguage{sdf}{%
  numbers=none,
  morekeywords={sorts,context-free,syntax,==,=,+,-,left,cons,prefer,avoid,bracket},
  columns=flexible,
  morestring=[b]",
  basicstyle=\footnotesize\mdseries,
  literate={->}{{\,\,$\to$\,\,}}1
}

\lstdefinelanguage{pp}{%
  numbers=none,
  morestring=[b]",
  stringstyle=\footnotesize\tt,
  literate={SPACE}{{\ }}1 {EPSILON}{{$\varepsilon$\,\,}}1 {*}{{$^\star$}}1 {+}{{$^+$}}1 {?}{{$?$}}1,
  keywordstyle=\normalfont\footnotesize\bfseries,
  morekeywords={abridge,addV,addH,anonymize,appear,chain,define,deanonymize,designate,detour,deyaccify,disappear,distribute,downgrade,dump,eliminate,extract,factor,fold,horizontal,inject,inline,introduce,lassoc,massage,narrow,permute,project,rassoc,redefine,renameN,renameL,renameT,renameS,replace,reroot,strip,terminalize,unchain,undefine,unfold,unite,unterminalize,upgrade,vertical,widen,yaccify,one,of,import,equate,unlabel,abstractize,concretize,removeV,removeH},
  columns=fullflexible,
  basicstyle=\footnotesize\it,
}

\lstdefinelanguage{tt}{%
  numbers=none,
  morekeywords={},
  morestring=[b]",
  columns=flexible,
  basicstyle=\footnotesize\tt,
}

\lstdefinelanguage{xbnf}{%
  numbers=none,
  morekeywords={},
  morestring=[b]",
  columns=flexible,
  basicstyle=\footnotesize\it,
  literate={EPSILON}{{$\varepsilon$\,\,}}1 {*}{{$^\star$\,\,}}1 {+}{{$^+$\,\,}}1 {?}{{$\!\!\! ?$}}1,
  morekeywords={sequence,abridge,add,anonymize,appear,chain,define,deanonymize,designate,detour,deyaccify,disappear,distribute,downgrade,dump,eliminate,extract,factor,fold,horizontal,inject,inline,introduce,lassoc,massage,narrow,permute,project,rassoc,redefine,remove,rename,replace,reroot,strip,terminalize,unchain,undefine,unfold,unite,unterminalize,upgrade,vertical,widen,yaccify},keywordstyle=\rmfamily\bfseries,
}

\lstdefinelanguage{prolog}{%
  numbers=left,
  morekeywords={},
  morestring=[b]',
  columns=flexible,
  commentstyle=\sffamily\slshape,
  morecomment=[l]{\%},
  literate={:-}{{\,$\Leftarrow$\,\,}}1 {-->}{{$\to$\,}}1
}

\lstdefinelanguage{xml}{%
  numbers=none,
  morekeywords={},
  sensitive=true,
  morecomment=[s]{<!--}{-->},
  morestring=[b]",
  stringstyle=,
  basicstyle=\small\tt,
  showspaces=false,
}

\newtheorem{crule}{Rule}

\newcommand{\ruleref}[1]{\hyperref[#1]{Rule \ref*{#1}}}
\newcommand{\exampleref}[1]{\hyperref[#1]{Example \ref*{#1}}}
\newcommand{\jlsquote}[1]{``\emph{#1}''}
\newcommand{\numberOfProductions}{Productions}
\newcommand{\numberOfNonterminals}{Nonterminals}
\newcommand{\myindent}{\ \ \ $\circ$\ }
\newcommand{\numberOfTops}{Tops}
\newcommand{\numberOfBottoms}{Bottoms}
\newcommand{\javaNumberOfRefactors}[8]{\myindent Semantics-preserving&#1&#2&#3&#4&#5&#6&#7&#8\\}
\newcommand{\javaNumberOfGeneralises}[8]{\myindent Semantics-increasing/-decreasing&#1&#2&#3&#4&#5&#6&#7&#8\\}
\newcommand{\javaNumberOfRevisings}[8]{\myindent Semantics-revising&#1&#2&#3&#4&#5&#6&#7&#8\\}
\newcommand{\javaNumberOfLines}[8]{Number of lines&#1&#2&#3&#4&#5&#6&#7&#8\\}
\newcommand{\javaNumberOfSteps}[8]{}
\newcommand{\javaNumberOfIssues}[8]{}
\newcommand{\javaNumberOfTransformations}[8]{Number of transformations&#1&#2&#3&#4&#5&#6&#7&#8\\}
\newcommand{\xbgfNumber}[1]{\myindent \emph{#1}}
\newcommand{\xml}[1]{\lstinline[language=xml]{#1}}
\newcommand{\noskip}{\topsep0pt \parskip0pt \partopsep0pt}
\newcommand{\tokenAlNum}{Alphanumeric}
\newcommand{\tokenBar}{$|$}
\newcommand{\tokenMeta}{\{,\},[,],(,)}
\newcommand{\tokenOther}{otherwise}
\newcommand{\javaIssuesPostX}[8]{}
\newcommand{\javaIssuesCorrect}[8]{}
\newcommand{\javaIssuesExtend}[8]{}
\newcommand{\javaIssuesPermit}[8]{}

\newcommand{\javaEarly}[8]{Convergence preparation&&&&&&&\\phase (\S\ref{X:preparation})&#1&#2&#3&#4&#5&#6&#7&#8\\}
\newcommand{\javaEarlyKnownBugs}[8]{\myindent Known bugs&#1&#2&#3&#4&#5&#6&#7&#8\\}
\newcommand{\javaEarlyPostExtraction}[8]{\myindent Post-extraction&#1&#2&#3&#4&#5&#6&#7&#8\\}
\newcommand{\javaEarlyInitialCorrection}[8]{\myindent Initial correction&#1&#2&#3&#4&#5&#6&#7&#8\\}
\newcommand{\javaFinal}[8]{Resolution phase&#1&#2&#3&#4&#5&#6&#7&#8\\}
\newcommand{\javaFinalCorrection}[8]{\myindent Correction (\S\ref{X:correction})&#1&#2&#3&#4&#5&#6&#7&#8\\}
\newcommand{\javaFinalRelaxation}[8]{\myindent Relaxation (\S\ref{X:relaxation})&#1&#2&#3&#4&#5&#6&#7&#8\\}
\newcommand{\javaFinalExtension}[8]{\myindent Extension (\S\ref{X:extension})&#1&#2&#3&#4&#5&#6&#7&#8\\}

\newcommand{\emphsub}[1]{\subsection*{\textbf{#1}}}	
\newcommand{\emphsubsub}[1]{\subsubsection{\textbf{#1}}}
\newcommand{\emphsec}[1]{\subsection{\textbf{#1}}}

\usepackage{hyperref}
\begin{document}\sloppy

\title{Recovering Grammar Relationships\\
	for the Java Language Specification}


\author{Ralf L\"ammel         \and
        Vadim Zaytsev}


\institute{R. L\"ammel \at
              Software Languages Team \\
              The University of Koblenz-Landau\\
              Germany\\
              \email{laemmel@uni-koblenz.de}
           \and
           V. Zaytsev \at
		              Software Languages Team \\
		              The University of Koblenz-Landau\\
		              Germany\\
		              \email{zaytsev@uni-koblenz.de}}

\date{Received: date / Accepted: date}

\maketitle

\begin{abstract}
  Grammar convergence is a method that helps discovering relationships
  between different grammars of the same language or different language
  versions. The key element of the method is the operational,
  transformation-based representation of those relationships. Given
  input grammars for convergence, they are transformed until they are
  structurally equal. The transformations are composed from
  primitive operators; properties of these operators and the composed
  chains provide quantitative and qualitative insight into the
  relationships between the grammars at hand.

  We describe a refined method for grammar convergence, and we use it
  in a major study, where we recover the relationships between all the
  grammars that occur in the different versions of the Java Language
  Specification (JLS). The relationships are represented as grammar
  transformation chains that capture all accidental or intended
  differences between the JLS grammars. This method is mechanized and
  driven by nominal and structural differences between pairs of
  grammars that are subject to asymmetric, binary convergence
  steps. 

  We present the underlying operator suite for grammar transformation
  in detail, and we illustrate the suite with many examples of
  transformations on the JLS grammars. We also describe the extraction
  effort, which was needed to make the JLS grammars amenable to
  automated processing. We include substantial metadata about the
  convergence process for the JLS so that the effort becomes
  reproducible and transparent.
\end{abstract}


%
\section{Introduction}

Overall, this paper is concerned with \textbf{the problem of grammar
  consistency checking}. Many software languages (and programming
languages, in particular) are described simultaneously by multiple
grammars that are found in different software artifacts. For instance,
one grammar may reside in a language specification; another grammar
may be encoded in a parser specification; yet another grammar may be
present in an XML schema for tool-independent data exchange. Ideally,
one would want to reliably establish and continuously maintain that
all co-existing (potentially embedded) grammars describe the same
intended language. Without such guarantee, grammar inconsistencies may
go unnoticed, and grammar-based software artifacts may get
brittle. Some existing ad-hoc or brute-force methods partially address
this problem, but ultimately grammar consistency checking is an open
software engineering problem without a satisfying best practice. A
good example is the Java Language Specification
\citep[JLS;][]{JLS1,JLS2,JLS3}, which is the target of the present
paper. The JLS is a critical specification in the software industry,
yet it contains substantial inconsistencies.

Let us sketch \textbf{the obstacles for grammar consistency
  checking.} Consider the problem of establishing or maintaining that
some given BNFs (i.e., grammars) describe the same language. An
automated solution is constrained by the formal undecidability of
grammar equivalence. Such a formal limit is certainly part of the
problem that there is no best practice for grammar consistency
checking. Obviously, the problem becomes even more challenging once we
consider the practical situation of grammars of many different forms:
BNFs, parser descriptions, XML schemata, software models, etc. Such
variation implies impedance mismatches. As a result, it may be hard to
mentally or automatically map one grammar to the other. The present
paper describes a method that addresses those obstacles effectively.

In essence, \textbf{grammar consistency checking deals with grammar
  differences} in a systematic manner. Grammars for the same language
may be different for various, practically viable reasons.  For
instance, grammars may be tailored for a certain purpose or quality
such as ``readability''. (In the present paper, we deal with ``more
readable'' vs. ``more implementable'' grammars for the Java language.)
Some grammars may have been designed independently of one another, and
hence they are likely to be vastly different in the sense of
structural equality of the grammar specifications. Other grammars may
have been affected heavily by compromises required by implementation
technologies (e.g., parsing techniques), or data models (e.g., XML
Schema as opposed to BNF). To summarize, in practice, there are many
intended, accidental, idiosyncratic, superficial, and substantial
differences between co-existing grammars of a language.

In fact, we need a generalized form of grammar consistency checking
that also account for \textbf{versatile grammar relationships due to
  software and language evolution}. Both, software languages as such
(e.g., in the form of language documentation) and grammar-based
software artifacts (e.g., compilers, source code analysis tools, IDEs)
are subject to possibly independent evolution. The grammars of
different versions are not even intended to describe the same
language, but one would still want to understand their relative
correspondence in terms of a delta between these versions. As a
result, there are even more grammars to be checked for
consistency. Also, we are no longer restricted to plain grammar
equivalence, but language extensions, restrictions, or revisions would
need to be captured and checked.

Another related challenge of language evolution is migration of data
(programs, words, etc.) across versions. We do not discuss this
challenge in the present paper, even though the underlying method may
be potentially useful in such a context.

In \citet{LaemmelZ08}, we have begun to address the fundamental
problem of grammar diversity by initiating a method for
\textbf{grammar convergence}. This method combines \emph{grammar
  extraction} (to obtain raw grammars from artifacts and represent
them uniformly), \emph{grammar comparison} (to determine nominal and
structural differences between given grammars), and \emph{grammar
  transformation} (to represent the relationships between given
grammars by transformations that make the grammars structurally
equal). Grammar convergence is another method of grammar
engineering---as such, it is a companion of grammar recovery,
adaptation, and inference. The specific property of convergence is
that it genuinely takes several grammars as input---as opposed to any
process that starts from a single grammar.

In the present paper, we describe \textbf{the JLS study}---a major
study for grammar convergence for the Java language. We also deliver
\textbf{a refined method for grammar convergence with improved
  scalability and reproducibility}.\footnote{An earlier and
  abbreviated account on this work has been published in the
  Proceedings of Ninth IEEE International Working Conference on Source
  Code Analysis and Manipulation, SCAM 2009, pp.\ 178--186.} The study in this paper
concerns the 3 different versions of the Java Language Specification
\citep[JLS;][]{JLS1,JLS2,JLS3}. Each of the 3 JLS versions contains 2
grammars: one grammar is said to be optimized for readability, and the
other one is intended as a basis for implementation.

Let us briefly discuss the JLS situation. One would expect that the
different grammars per version are essentially equivalent in terms of
the generated language. As a concession to practicality (i.e.,
implementability, in particular), one grammar may be more permissive
than the other. One would also expect that the grammars for the
different versions generate languages that engage in an inclusion
ordering because of the backwards-compatible evolution of the Java
language. \emph{Those expected relationships of (liberal) equivalence
  and inclusion ordering are significantly violated by the JLS
  grammars, as our study shows.}

The JLS is critical to the Java platform---it is a foundation for
compilers, code generators, pretty-printers, IDEs, source code analysis
and manipulation tools, and other grammarware for the Java language. The
JLS is the authoritative specification of Java. Hence, there is a
strong incentive for an unambiguous, consistent and understandable set
of JLS documents. Still, our study discovers substantial
inconsistencies with the help of grammar convergence.

Our work is in no way restricted to the JLS.  We notice \textbf{a
  broader impact on language standardization and engineering}. Based
on the major JLS study of the present paper, previous work on grammar
recovery, and general trends in software language engineering, we
contend that grammar convergence improves the state of the art in
creation, maintenance and evolution of language
documentation. Ideally, we would hope for standardization bodies and
language documenters to incorporate grammar convergence into their
methodology. For instance, it would be clearly desirable for Oracle to
abandon manual grammar editing in the next version of Java and the
JLS. We refer to \citet{NeedsGrammarware} for a proposal, in fact, an
ISO document, that hints at the application of grammar engineering
techniques such as grammar convergence in the context of creating,
maintaining, or evolving language documents. Realistically, though, it
will be difficult to replace current ad-hoc techniques of dealing with
multiple grammars in language documents and otherwise. We will discuss
some of the limitations of the current grammar convergence method in
the conclusion. There is the particular issue of adoption: it would
take Oracle, ISO, and other such organizations substantial effort to
incorporate additional methods and tools into their processes, and to
adjust existing documents. Overall, grammar convergence is still an
emerging method.

\subsection*{\textbf{Contributions}}

The motivation of our work and its significance is not limited to the
mere discovery of bugs in the Java standard or in any other set of
grammars for that matter. (In fact, some JLS bugs have been
discovered, time and again, by means of informal grammar inspection or
other brute-force methods.) The significance of our work is amplified
by two arguments. First, we provide a simple and mechanized process
for discovering accidental or intended differences between
grammars. Second, we are able to represent the differences in a
precise, operational and accessible manner---by means of grammar
transformations.

Here is an itemized summary of the contributions of this work:

\begin{enumerate}

\item We have recovered nontrivial relationships between grammars of
  industrial size. (That is, we show that the grammars are equivalent
  modulo well-defined transformations.) 

\medskip

\item We have designed a mechanized, measurable and reproducible
  process for grammar convergence. Compared to the initial work on
  grammar convergence \citep{LaemmelZ08}, the process consists of
  well-defined phases and its progress can be effectively tracked in
  terms of the numbers of nominal and structural differences between
  the grammars at hand.

\medskip

\item We have worked out a comprehensive operator suite for grammar
  transformation driven by the scale of the present JLS study. The suite
  substantially improves on prior art.

\medskip

\item The complete JLS effort (including all the involved sources,
  transformations, results, and tools) is publicly available through
  SourceForge.\footnote{\url{http://slps.sf.net/}; see
    \href{http://slps.svn.sourceforge.net/viewvc/slps/topics/java/lci/}{\texttt{topics/java/lci}}
    in particular.}

\end{enumerate}

\subsection*{\textbf{Roadmap}}

\S\ref{S:convergence} gives an overview on grammar convergence method,
it prepare the application of the method to the JLS, and it describes
phases of a refined process of convergence that we extracted from the
reported JLS study. \S\ref{S:extraction} describes the extraction
phase of grammar convergence for the JLS. \S\ref{S:transformation}
describes an operator suite for grammar transformation, and applies it
to the JLS. \S\ref{S:measures} provides a postmortem for the reported
JLS study.  \S\ref{S:related} discusses related work. \S\ref{S:concl}
concludes the paper.

\section{Grammar convergence}
\label{S:convergence}

The central idea of grammar convergence~\citep{LaemmelZ08} is to
extract grammars from diverse software artifacts, and to discover and
represent the relationships between the grammars by chains of
transformation steps that make the grammars structurally equal. In
this section, we will describe the method in detail and prepare its
application to the JLS. The method relies on the following core
ingredients:

\begin{itemize}

\item A unified \emph{grammar format} that effectively supports
  abstraction from specialities or idiosyncrasies of the grammars as
  they occur in software artifacts in practice.

\item A \emph{grammar extractor} for each kind of artifact.  (In the
  present JLS study, we had to extract grammars from the JLS
  documents, which are available in HTML and PDF.)

\item A \emph{grammar comparator} that determines and reports grammar
  differences in the sense of deviations from structural equality.

\item A framework for automated \emph{grammar transformation} that can
  be used to refactor, or to otherwise more liberally edit grammars
  until they become structurally equal.
\end{itemize}


\emphsec{Grammar comparison} 
\label{S:comp-trafo}

\begin{figure}
	\input{examples/class-declaration-compare.tex}
\caption{Two similar grammar excerpts from different versions of the JLS.
	The second excerpt involves two more nonterminals than the first excerpt:
	\emph{NormalClassDeclaration}, which looks similar to the nonterminal
	from the first grammar, and \emph{EnumDeclaration}, which is completely
	new. Hence, we speak of two nominal differences (two nonterminals in
	\emph{read3} that do not match \emph{read2}), and of two structural
	differences (two unmatched branches in \emph{ClassDeclaration}).}
\label{F:compare1}
\end{figure}

The grammar comparator is used to discover grammar differences, and
thereby, to help with drafting transformations in a stepwise
manner. We distinguish nominal vs.\ structural grammar differences. We
face a \emph{nominal difference} when a nonterminal is defined or
referenced in one of the grammars but not in the other. We face a
\emph{structural difference} when the definitions of a shared
nonterminal differ for two given grammars. Some of the nominal
differences will be eliminated by a simple renaming, while others will
disappear gradually when dealing with structural differences that
involve folding/unfolding.

In order to give better results, we assume that grammar comparison operates on
a slightly normalized grammar format. The assumed, straightforward
normalization rules are presented in \autoref{A:normalize}.

Let us consider a simple example, without paying attention yet to the specific
grammar notation and transformation operators. For instance, consider the two
grammar excerpts from the ``more readable'' grammars of JLS2 and JLS3
(\emph{read2} and \emph{read3} from now on) in \autoref{F:compare1}.
Conceptually, the grammars are different in the following manner. The
\emph{read3} grammar covers additional syntax for enumeration declarations; it
also uses an auxiliary nonterminal \emph{NormalClassDeclaration} for the
class-declaration syntax that is declared directly by \emph{ClassDeclaration}
in the \emph{read2} grammar. The comparator reports four differences that are
directly related to these observations:

\medskip

\noindent
\begin{boxedminipage}{\hsize}
\begin{itemize}
\item Nominal differences: 
\begin{itemize}
\item read2: nonterminal \emph{NormalClassDeclaration} missing.
\item read2: nonterminal \emph{EnumDeclaration} missing.
\end{itemize}
\item Structural differences:
\begin{itemize}
\item Nonterminal \emph{ClassDeclaration}: no matching alternatives\\
 (counts as 2 because the definitions have a maximum of 2 alternatives).
\end{itemize}
\end{itemize}
\end{boxedminipage}

\medskip

Arguably, these differences should help the grammar engineer who will typically
try to find definitions for missing nonterminals by extracting their inlined
counterparts. The counterpart for \emph{NormalClassDeclaration} is relatively
obvious because of the combination of a nonterminal that is entirely missing in
one grammar while it occurs in a structural different and unmatched alternative in
the other grammar.

\emphsec{Grammar transformation}

\begin{figure}
	\input{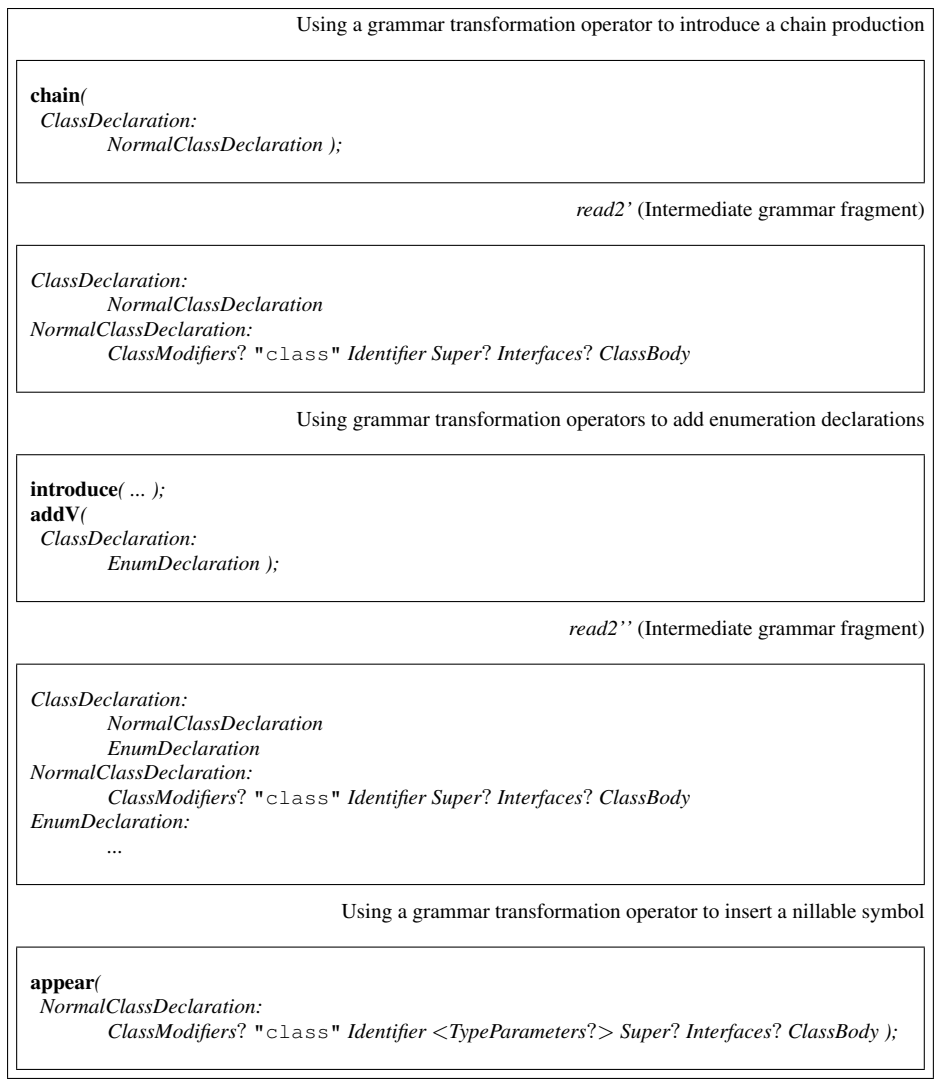}
\caption{Transforming the grammar and proving $(\textbf{chain} \circ
\textbf{introduce} \circ \textbf{addV} \circ \textbf{appear})\textrm{-equality}$.}
\label{F:transform1}
\end{figure}

Since the goal of grammar convergence is to relate all sources to each
other, the relationships between grammars will be represented as
grammar transformations. We say that grammars $g_1$ and $g_2$ are
$f\textrm{-equal}$, if $f(g_1)=g_2$ (where ``$=$'' refers to
structural equality on grammars, and $f$ denotes the meaning of a
grammar transformation). When $f$ is a refactoring (i.e., a
semantics-preserving transformation), then $f\textrm{-equality}$
coincides with grammar equivalence. If $f$ is a semantics-increasing
(-decreasing) transformation, then we have shown an inclusion ordering
for the languages generated by the two grammars.

We use the terms ``semantics-preserving'', ``-increasing'' and
``-decreasing'' in the formal sense of the language generated by a
grammar. Clearly, the composition of (sufficiently expressive)
increasing and decreasing operators allows us to relate arbitrary
grammars, in principle. Hence, more restrictions are needed for
accumulating reasonable grammar relationships, as we will discuss
below. We also mention that there is a rare need for operators that
are neither semantics-increasing nor -decreasing. In this case, we
speak of a semantics-revising operator. Consider, for example, an
unconstrained \textbf{replace} operator for expressions in grammar
productions that may be needed if we face conflicting definitions of a
nonterminal in two given grammars.

The baseline scenario for grammar transformation in the context is convergence
is as follows. Given are two grammars: $g_1$ and $g_2$. The goal is to find $f$
such that $g_1$ and $g_2$ are $f\textrm{-equal}$. In this case, one has to
gradually aggregate $f$ by addressing the various differences reported by the
comparator. In our current implementation of grammar comparison, we do not make
any effort to propose any transformation operators to the user, but this is
clearly desirable and possible.

In JLS, given the differences reported by the comparator and presented
in the previous section, the grammar engineer authors an
transformation to add an extra chain production for
\emph{NormalClassDeclaration}. This transformation and a few
subsequent ones as well as all intermediate results are listed in
\autoref{F:transform1}. 

The idea is now that such compare/transformation steps are repeated.
Hence, we compare the intermediate result, as obtained above, with the
grammar \emph{read3}. It is clear that the nominal difference for
\emph{NormalClassDeclaration} has been eliminated. The comparator
reports the three following differences:

\medskip

\noindent
\begin{boxedminipage}{\hsize}
\begin{itemize}
\item Nominal difference:
\begin{itemize}
\item read2': nonterminal \emph{EnumDeclaration} missing.
\end{itemize}
\item Structural difference: nonterminal \emph{NormalClassDeclaration}
\begin{itemize}
	\item read2': \lstinline[language=pp]|ClassModifiers? SPACE "class" SPACE Identifier SPACE Super? SPACE Interfaces? SPACE ClassBody|
	\item read3: \lstinline[language=pp]!ClassModifiers? SPACE "class" SPACE Identifier SPACE TypeParameters? SPACE Super? SPACE Interfaces? SPACE ClassBody!
\end{itemize}
\item Structural difference: nonterminal \emph{ClassDeclaration}
\begin{itemize}
\item Unmatched alternatives of read2': none
\item Unmatched alternatives of read3: \lstinline[language=pp]{EnumDeclaration}
\end{itemize}
\end{itemize}
\end{boxedminipage}

\medskip

We see that enumerations are missing entirely from \emph{read2'}, and hence a
definition has to be introduced, and a corresponding alternative has to be
added to \emph{ClassDeclaration}. Once we are done, the result is again
compared to \emph{read3}:

\medskip

\noindent
\begin{boxedminipage}{\hsize}
\begin{itemize}
\item Structural difference: nonterminal \emph{NormalClassDeclaration}
\begin{itemize}
	\item read2'{}': \lstinline[language=pp]|ClassModifiers? SPACE "class" SPACE Identifier SPACE Super? SPACE Interfaces? SPACE ClassBody|
	\item read3: \lstinline[language=pp]!ClassModifiers? SPACE "class" SPACE Identifier SPACE TypeParameters? SPACE Super? SPACE Interfaces? SPACE ClassBody!
\end{itemize}
\end{itemize}
\end{boxedminipage}

\medskip

Again, this difference is suggestive. Obviously, the definition of
\emph{NormalClassDeclaration} according to \emph{read2''} does not
cover the full generality of the construct, as it occurs in
\emph{read3}. The structural position for the type parameters of a
class has to be added. (This has to do with Java generics which were
added in the 3rd edition of the JLS.) There is a designated
transformation operator that makes new components \textbf{appear}
(such as type parameters) in existing productions; the newly inserted
part is marked on \autoref{F:transform1} with angle brackets. This is
a downward-compatible change since type parameters are optional. Once
these small transformations have been completed, all the discussed
differences are resolved, and the comparator attests structural
equality.


\emphsec{Convergence graphs} 

Grammar convergence always starts from the grammars that were extracted
from the given software artifacts, to which we refer as \emph{source
grammars} or \emph{sources} subsequently. In the present JLS study, we
face 6 sources; we use \emph{read1}--\emph{read3} to refer to the
``more readable'' grammars, and \emph{impl1}--\emph{impl3} to refer to
the ``more implementable'' grammars. It is reasonable to relate
grammars through an additional grammar of which we think as the common
denominator of the original grammars. We refer to such additional
grammars as \emph{targets}. The ``distance'' between source and target
grammars may differ. In fact, it is not unusual, that one
source---modulo minor transformations only---serves as common
denominator.

The idea of the common denominator can be generalized such that we
actually devise a \emph{directed acyclic graph} with grammars as the
nodes and transformations as the edges. In the trivial case with each
target being a result of transforming two sources or targets, we will
have a \emph{binary tree}. The root of such a tree (the final target)
is the common denominator of all grammars, but there may be additional
intermediate targets that already serve as common denominators for some
of the grammars. (We use arrows to express the direction of the
transformation, and hence the trees appear inverted, when compared to
common sense of drawing trees.) The source grammars are the leaves of
such a tree.

\begin{figure}
\begin{boxedminipage}{\hsize}
\begin{tabular}{lr}
\includegraphics[width=.65\textwidth]{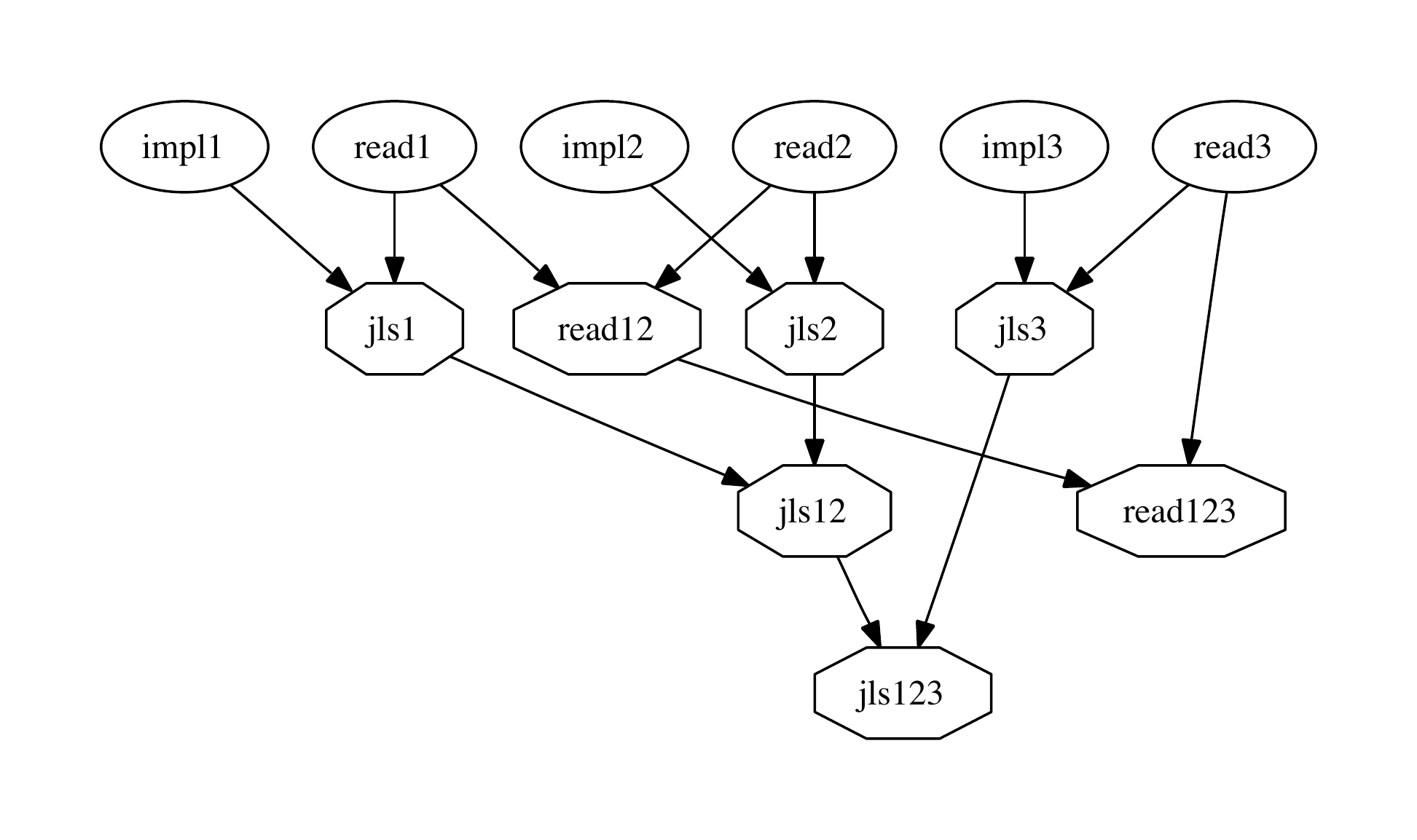}&
\raisebox{10em}{\begin{tabular}{l|l}
	\emph{impl1}	& JLS1, \S19\\
	\emph{read1}	& JLS1, \S\S4.1--15.27\\
	\emph{impl2}	& JLS2, \S18\\
	\emph{read2}	& JLS2, \S\S4.1--15.28\\
	\emph{impl3}	& JLS3, \S18\\
	\emph{read3}	& JLS3, \S\S4.1--15.28
\end{tabular}}
\end{tabular}
\end{boxedminipage}

\caption{The convergence graph for the JLS grammars consists of two
binary trees with shared leaves. The \emph{nodes} in the figure are
grammars where the leaves correspond to the original JLS grammars and
the other nodes are derived. The \emph{directed edges} denote grammar
transformation chains. We use a (cascaded) binary tree here, i.e., each
forking node is derived from two grammars. The \emph{implX} leaves are
``implementable'' grammars, the \emph{readX} ones are ``readable''.}
\label{F:tree}
\end{figure}

\autoref{F:tree} shows the ``convergence tree'' for the present JLS
case study. The original grammars from the JLS documents are located
at the top. The tree states that the two grammars per JLS version are
``converged to'' a common denominator (see the nodes \emph{jls1--3} in
the figure), and all three versions are further ``converged'' to
account for inter-version differences---the extensions to the Java
language in particular (see the nodes \emph{jls12} and \emph{jls123}
as well as \emph{read12} and \emph{read123} in the figure). For the
JLS we use a binary tree, which means that we always limit the focus
to two grammars, and hence a cascade is needed, if more than two
grammars need to be converged.

When deriving \emph{jls1--3}, we favor the ``more implementable''
grammar as the target of convergence, i.e., as the common
denominator---except that some corrections may need to be applied, or
some minimum restructuring is applied for the sake a more favorable
grammar structure. This preference reflects the general rule that an
implementation-oriented artifact should be derived from a
design-oriented artifact---rather than the other way
around. Incidentally, this direction is also easier to handle by the
available transformation operators.

When relating the different JLS versions, we adopt the redundant
approach to relate the common denominators \emph{jls1--3} in one
cascade (see the nodes \emph{jls12} and \emph{jls123}), but also the
readable grammars \emph{read1--3} in another cascade (see the nodes
\emph{read12} and \emph{read123}) as a sort of sanity check. It turns
out that \emph{read1--3} are structurally quite similar, and
accordingly, the additional cascade requires little effort.


\emphsec{Convergence process} 
\label{S:process}

As we were discussing grammar comparison and transformation, we
already alluded to a basic compare/transform cycle---this cycle is
indeed the spline of the convergence process. We identify
\emph{phases} for the convergence process in order to impose more
structure and discipline onto the process. These convergence phases
assume asymmetric, binary convergence trees where one of the two
grammars is favored as (near-to) common denominator---as we discussed
above. There are five consecutive convergence phases: the initial
extraction phase involves a mapping from an external grammar format
and is therefore implemented as a standalone tool in our
infrastructure; the other four convergence phases are directly
concerned with transformation.

\begin{description}

\item[\textbf{Extraction:}] A starting point for grammar extraction is
  always a set of real grammar artifacts. A mapping is required for
  each kind of artifact so that grammar knowledge can be extracted and
  represented in a uniform grammar format.  (In in the case of our
  infrastructure, we use BGF---a BNF-like Grammar Format.) Each
  extractor may implement particular design decisions in terms of any
  normalization or abstraction to be performed along with
  extraction. Once extraction is completed, a (possibly incorrect or
  not fully interconnected) grammar is ready for transformation.

\item[\textbf{Convergence preparation:}] This convergence phase
  involves correcting immediately obvious or a priori known errors in
  the given sources. These corrections are represented as grammar
  transformations so that they can be easily revisited or re-applied
  in the case when the extractor is modified or the source changes. In
  the JLS case, we incorporated an available bug list at this
  stage\footnote{There are various accounts that have identified or
    fixed bugs in the JLS grammars or, in fact, in grammars that were
    derived from the JLS in some manner. We refer to the work of
    Richard Bosworth as a particularly operational account; it is a
    clear list of bugs which was also endorsed by Oracle:
    \url{http://www.cmis.brighton.ac.uk/staff/rnb/bosware/javaSyntax/syntaxV2.html}.
    We refer to this list as ``known bugs'' in our process.}. Some
  inaccuracies caused by representation anomalies in the HTML input
  were also resolved a this stage. Further, we added some missing
  definitions the lack of which was discovered through an early
  inspection; see the discussion of bottom nonterminals in
  \S\ref{S:jls-metrics}.

\item[\textbf{Nominal matching:}] We perform asymmetric
  compare/transform steps. That is, the non-favored grammar is
  compared with the (prepared) favored grammar, which is the baseline
  for the (intermediate) target of convergence. The
  objective of this convergence phase is to align the syntactic
  categories of the grammars in terms of their nonterminals. The
  nominal differences, as identified by comparison, guide the grammar
  engineer in drafting transformations for renaming as well as
  extraction and inlining such that the transformations immediately
  reduce the number of nominal differences. It is important to notice
  that we restrict ourselves to operators for renaming, inlining, and
  extraction. These operators convey our intuition of (initial) nominal
  alignment. We make these assumptions:

	\begin{itemize}

	\item
	  \emph{When a nonterminal occurs in both grammars, then it models the
	  same syntactic category (conceptually).} If the assumption does not
	  hold, then this will become evident later through considerable
	  structural differences, which will trigger a renaming to resolve the
	  name clash. Such corrective renaming may be pushed back to the phase
	  of convergence preparation.

	\item
	  \emph{Any renaming for nonterminals serves the purpose of giving the
	  same name to the same syntactic category (in an conceptual sense).}
	  If a grammar engineer makes a mistake, then this will become
	  evident later, again, through considerable structural differences. In
	  this case, we assume that the grammar engineer returns to the name
	  matching phase to revise the incorrect match.
  
	\end{itemize}

\item[\textbf{Structural matching:}]
  We continue with asymmetric compare/transform steps. This convergence phase
  dominates the transformation effort; it aligns the definitions of the
  nonterminals in a structural sense. The structural differences, as identified
  by comparison, guide the grammar engineer in drafting transformations for
  refactoring such that they immediately reduce the number of structural
  differences. As we continue to limit ourselves to refactoring, the order of
  the individual transformations does not matter due to its commutativity. The
  grammar engineer can simply pick any applicable refactoring operator, but the
  firm requirement is that the number of structural and nominal differences
  declines, which is automatically verified by our infrastructure.

\item[\textbf{Resolution:}] This convergence phase consists of three
  kinds of steps, as discussed in more detail in
  \S\ref{S:transformation}: \emph{extension}, \emph{relaxation} and
  \emph{correction}. In the case of semantics-increasing operators, it
  is up to the grammar engineer to perform the
  classification. Semantics-decreasing operators serve correction on
  the grounds of a convention. That is, we assume a directed process
  of convergence where the grammars of extended (relaxed) languages
  are derived from the grammars of ``sublanguages''. However, if the
  grammars violate such an intended sublanguage relationship, then
  correction must be expressed through semantics-decreasing operators.

\end{description}

The correctness of the process relies on one assumption regarding the
limited use of non-semantics-preserving operators. In particular,
non-semantics-preserving operators should only be used, if the given
grammars are not equivalent. Making equivalent grammars non-equivalent
is clearly not desirable. Currently, we cannot verify this assumption,
and in fact, it is generally impossible because of undecidability of
grammar equivalence. However, a heuristic approach may be feasible,
and provides an interesting subject for future work. Even when the
given grammars are non-equivalent, we still need to limit the use of
non-semantics-preserving operators for correctness' sake. That is, we
should disallow zigzag transformations such that semantics-increasing
and -decreasing transformations partially cancel each other.

\begin{figure}[t!]
\begin{center}
\includegraphics[width=0.75\textwidth]{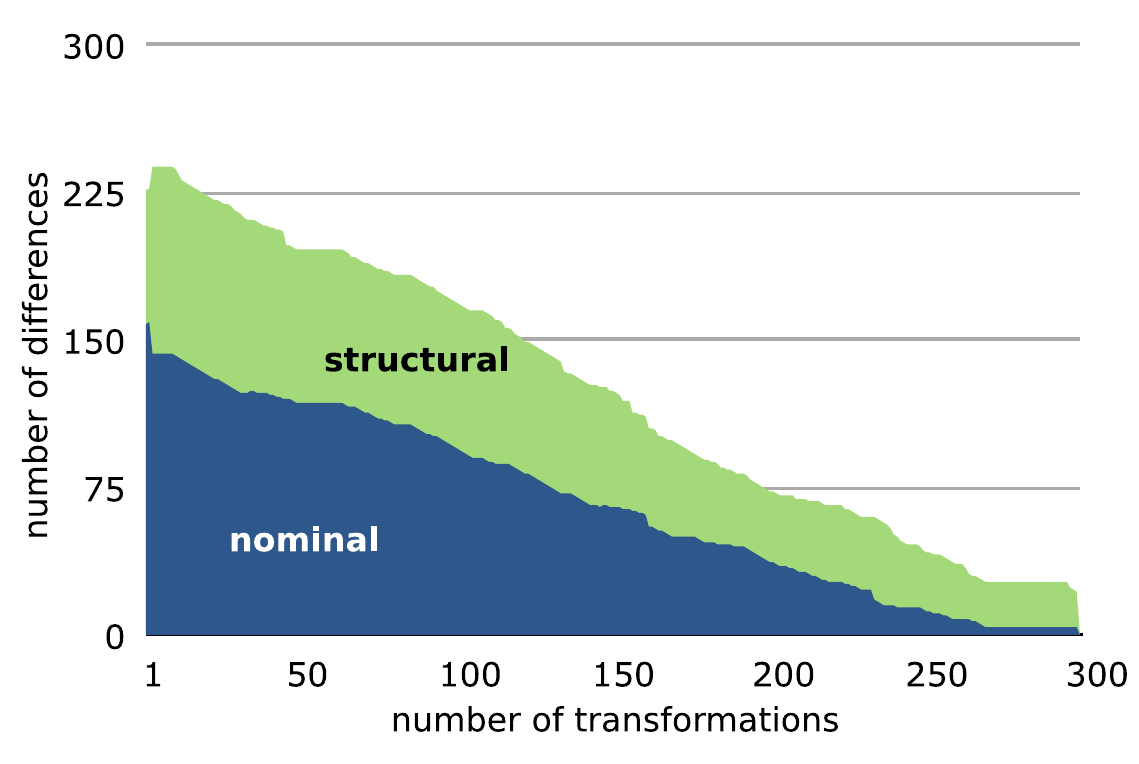}
\end{center}
\vspace{-42\in}
\caption{Difference reduction for \emph{read2} towards the convergence
  target \emph{jls2} in the convergence tree of \autoref{F:tree}.}
\label{F:workflow}
\end{figure}


We use the number of nominal and structural differences as means to
\textbf{track progress} of grammar convergence. Each unmatched nonterminal
symbol of either grammar counts as a nominal difference. For every nominally
matched nonterminal, we add the maximum number of unmatched alternatives (of
either grammar), if any, to the number of structural differences.

The main guiding principle for grammar convergence is to consistently
reduce the number of grammar differences throughout the two matching
convergence phases as well as the final resolution
phase. \autoref{F:workflow} illustrates this principle for one edge in
the convergence graph of the present JLS study. The figure also
visualizes that nominal differences tend to be resolved earlier than
structural differences.

Our transformation infrastructure is aware of the different phases of
convergence, and it checks the incremental reduction of differences at
runtime. As a concession to a simple design of the operator suite for
grammar transformations, restructuring steps may also slightly
increase structural differences as long as they are explicitly grouped
in ``transactions'' whose completion achieves reduction.

\section{Grammar extraction}
\label{S:extraction}

The previous section decomposed grammar convergence essentially into
grammar extraction and a compare/transform cycle. The present section
will focus on extraction, whereas the next section covers grammar
transformations to be used in the compare/transform cycle. (Here we
assume that our grammar comparison approach is currently trivial and
not worth a designated, detailed description.)

The central objective of grammar extraction is to map software
artifacts of a given kind (such as parser descriptions, language
documentation, or XML schemata) to the uniform grammar format that is
used in a convergence effort. Several engineering issues arise in this
context:
\begin{description}

\item[\textbf{Domain analysis.}]  One needs to understand the software
  artifacts at hand. One dimension of understanding may be based on
  metadata about the grammars at hand: version, style,
  completeness. Such information can be often obtained by inspecting
  the given grammar artifacts, e.g., language documents. Consider, for
  example, encountering of a left-recursive style. Awareness of this
  style is beneficial for the transformations to be devised
  eventually.

\item[\textbf{Source selection.}]  There may be multiple potential
  candidates (think of PDF vs.\ HTML for language documentation, or
  Java sources vs.\ byte code for object models). Hence, trade-offs
  regarding the simplicity, robustness, correctness, and completeness
  of extraction must be considered. Either one contender is chosen, or
  multiple options are explored in parallel, or a fallback is
  considered, if the contender fails, eventually.

\item[\textbf{Extractor implementation.}]  In our experience, most
  extractors are unique programs---they require particular programming
  techniques, specifically parsing techniques. Seeking the right
  implementation strategy is a matter of trial-and-error, but,
  ultimately, it is important to be able to describe a lucid
  implementation strategy so that one can have trust in the
  robustness, correctness, and completeness of extraction.

\item[\textbf{Metrics assessment.}]  One should evaluate the initial
  quality of the extracted grammar based on common grammar metrics for
  bottom nonterminals (``undefined nonterminals'') and top
  nonterminals (``unused nonterminals''). Such quality properties are
  helpful in guiding subsequent transformation efforts. Also, quality
  issues may indicate flaws in the extractor logic, and hence trigger
  reconsideration and revision.

\end{description}

\noindent
These issues will be addressed in the following text.


\emphsec{JLS domain analysis}
\label{S:jls-analysis}

We have already begun capturing JLS terminology, recall the notions of
``more readable'' and ``more implementable''. These notions are not
sharply defined, but one can think of, for example, \emph{left
  factoring} (to help with look ahead) as being used in the more
implementable grammars but not in the more readable grammars.  Let us
extract related characteristics of the grammars from the JLS documents
on a per-grammar basis:

\begin{description}

\item[\textbf{JLS1}] It is stated \citep[\S19]{JLS1} that the more
  implementable grammar has \jlsquote{been mechanically checked to
    insure that it is LALR(1)}. The correspondence between
  \emph{read1} and \emph{impl1} is briefly described by
  saying~\citep[\S2.3]{JLS1} that \emph{read1} is \jlsquote{very
    similar to} \emph{impl1} \jlsquote{but more readable}.

\item[\textbf{JLS2}] The second edition of the JLS~\citep[``Preface to
  the Second Edition'']{JLS2} \jlsquote{integrates all the changes
    made to the Java programming language since [...] the first
    edition in 1996. The bulk of these changes [...] revolve around
    the addition of nested type declarations.} The JLS1/2 grammars
  themselves are nowhere related explicitly. Upon cursory examination
  we came to conclude that \emph{read1} and \emph{read2} are
  strikingly similar (modulo the extensions to be expected), whereas
  surprisingly, \emph{impl1} and \emph{impl2} appeared as different
  developments. Also, the LALR(1) claim for \emph{impl1} is not
  matched by \emph{impl2} which does not list a grammar-class claim.
  However, \emph{impl2} is said~\citep[\S18]{JLS2}
  to be \jlsquote{the basis for the reference implementation}.

\item[\textbf{JLS3}] JLS3 extends JLS2 in numerous
  ways~\citep[Preface]{JLS3}: \jlsquote{Generics, annotations,
    asserts, autoboxing and unboxing, enum types, foreach loops,
    variable arity methods and static imports have all been added to
    the language}. Again, the JLS2/3 grammars themselves are nowhere
  related explicitly, and again, cursory examination suggests that
  \emph{read2} and \emph{read3} are strikingly similar (modulo the
  extensions to be expected). This time, \emph{impl2} and \emph{impl3}
  also bear strong resemblance. No definitive grammar-class claim is
  made, but an approximation thereof: \emph{impl3} is
  said~\citep[\S18]{JLS3} to be \jlsquote{not an LL(1) grammar, though
    [...] it minimizes the necessary look ahead.} Hence, \emph{impl3}
  has definitely departed from \emph{impl1} with its associated
  grammar class LALR(1).

\end{description}

\begin{table}[t!]\normalsize
\begin{center}
\begin{tabular}{l|c|c}
&\textbf{Grammar class}
&\textbf{Iteration style}
\\ \hline\hline
\emph{impl1} & LALR(1) & left-recursive \\ \hline
\emph{read1} & none & left-recursive \\ \hline
\emph{impl2} & unclear & EBNF metasymbols \\ \hline
\emph{read2} & none & left-recursive \\ \hline
\emph{impl3} & ``nearly'' LL(k) & EBNF metasymbols \\ \hline
\emph{read3} & none & left-recursive \\ \hline
\end{tabular}
\end{center}
\caption{Basic properties of the JLS grammars.}
\label{F:classes-and-iteration-style}
\end{table}

In addition to grammar class claims for the JLS grammars we have also
recorded iteration styles during cursory examination; see
\autoref{F:classes-and-iteration-style}. This data already clarifies
that we need to bridge the gap between different iteration styles
(which is relatively simple) but also different grammar classes (which
is more involved)---if we want to recover the relationships between
the different grammars by effective transformations.


\emphsec{JLS source selection}
\label{S:jls-source}

A JLS document is basically a structured text document with embedded
grammar sections. In fact, the ``more readable'' grammar is developed
throughout the document where the ``more implementable'' grammar is
given, \emph{en bloc}, in a late section---a de facto appendix.

The JLS is available electronically in HTML and PDF format. Neither of
these formats was designed with convenient access to the grammars in
mind. After some deliberation, we have opted for the HTML format
because parsing seemed relatively straightforward.

Obviously, the JLS grammars have been implemented by different parties
in various ways. For instance, there exist parser descriptions whose
authors have consulted the JLS. However, as a matter of principle,
none of these options was considered appropriate in the present JLS
study because we wanted to make sure to perform consistency checking
for the primary JLS as opposed to any derived artifact. Hence, we
started from the JLS (and its HMTL documents, in particular), even if
such a source selection required more effort than a path that reuses
third-party Java grammars.


\emphsec{JLS grammar notation}
\label{S:scrapping}

The extractor needs to identify grammar portions within general HTML
markup. The used grammar format slightly varies across the different
JLS grammars and versions; there are relevant formatting rules in
different documents and sections---in particular from
\citet[\S2.4]{JLS1}, \citet[\S2.4, \S18]{JLS2} and \citet[\S2.4,
\S18]{JLS3}.

\begin{figure}[p]
\input{examples/assumed.tex}
  \caption{Relevant expressiveness of the JLS grammar notation, given in a self-descriptive
    manner; for clarity, terminals are enclosed in double quotes as
    opposed to the use of markup; the markup-based form of optional symbols
    is also omitted.}
    \label{F:grammar}
\end{figure}

\begin{figure}[p]
\input{examples/bgf.tex}
  \caption{In this paper, we show grammar fragments in a pretty-printed format (as opposed to
  the markup-based source format): nonterminals are in italic type;
  terminals are enclosed in double quotes; operators ``?'', ``*''
  and ``+'' serve for optionality and lists; 
  elisions are shown as ``...''.}
    \label{F:bgf}
\end{figure}

Grammar fragments are hosted by \xml{<pre>...</pre>} blocks in the JLS
documents. According to \citet[\S2.4]{JLS1,JLS2,JLS3}: terminal symbols
are shown in fixed font (as in \xml{<code>class</code>}); nonterminal
symbols are shown in italic type (as in \xml{<i>Expression</i>}); a
subscripted suffix ``opt'' indicates an optional symbol (as in
\xml{Expression<sub>opt</sub>}); alternatives start in a new line and
they are indented; ``one of'' marks a top-level choice with atomic
branches. (We have also observed that nonterminals are expected to be
alphanumeric and start in upper case.) Further notation and
expressiveness is described in~\citet[\S18]{JLS2,JLS3}: $[x]$ denotes
zero or one occurrences of $x$; $\{x\}$ denotes zero or more
occurrences of $x$; $x_1|\cdots|x_n$ forms a choice over the
$x_i$. The JLS documents consistently suffice with ``*'' lists (zero
or more occurrences); there are no uses of ``+'' lists. Refer to
\autoref{F:grammar} for a summary of the assumed source grammar notation.
Refer to \autoref{F:bgf} for a summary of the notation we use for the
examples in this paper. All examples presented here were obtained
from their XML (BGF) form in an automated generative manner as in \citet{GDK}.

We should also mention line continuation; it allows to spread one
alternative over several lines~\citep[\S2.4]{JLS3}: \jlsquote{A very
long right-hand side may be continued on a second line by
substantially indenting this second line}.
In our notation we double the indentation for every continued line.

\medskip

\begin{example}
\label{X:grammar}
A grammar fragment as of~\citet[\S4.2]{JLS2}:

\begin{lstlisting}[language=tt]
<i>NumericType:
        IntegralType
        FloatingPointType

IntegralType: one of</i>
        <code>byte short int long char
</code>
\end{lstlisting}

It should be parsed as:

\medskip

\noindent
\begin{boxedminipage}{0.5\textwidth}
\begin{lstlisting}[language=pp]
NumericType:
        IntegralType
        FloatingPointType

IntegralType:
        "byte"
        "short"
        "int"
        "long"
        "char"
\end{lstlisting}
\end{boxedminipage}

\medskip

The fragment illustrates two different kinds of ``choices'', i.e.,
multiplicity of vertical alternatives, and ``one of'' choices. (The
third form, which is based on ``$|$'', is not illustrated.) The
fragment also clarifies that markup tags are used rather
liberally. The ``nonterminal'' tag (i.e., \xml{<i>...</i>}) spans more
than one production. The terminal tag (i.e., \xml{<code>...</code>})
spans several terminals and the closing tags ends up on a new line.
\end{example}

\medskip

\begin{example}
	Not only the indentation is incorrect in the following fragment,
	it is also the only place where the subscript ``opt'' is capitalized
	\citep[\S4.5.1]{JLS3}:

\begin{lstlisting}[language=tt]
Wildcard:
? WildcardBounds<sub>Opt</sub>
\end{lstlisting}
\end{example}


\emphsec{JLS extractor implementation}
\label{S:jls-extractor}

The tiny \autoref{X:grammar} is a good indication of the many irregularities that
are found in the HTML representation, such as volatile use of markup tags, liberal
indentation, duplicate definitions. We needed to design and implement a non-classic
grammar parser to extract and analyze the grammar segments of the documents and to
perform recovery. Our extractor therefore deals with the expected irregularities in
several phases.

\begin{itemize}
	\item Phase 1---Preprocessing: the tool takes an HTML formatted text
			and filters out all hypertext tags and indentation, extracting
			all possible information from them in the process.
	\item Phase 2---Error recovery: the recovery rules are applied until they are
			no longer applicable. There are rules for transforming a terminal
			symbol to a nonterminal symbol or the other way around,
			matching up parentheses, splitting/combining sibling symbols, etc.
	\item Phase 3---Removal of doubles: duplicate definitions are purged.
		This could not happen during earlier extraction phases because
		clones could differ in markup.
	\item Phase 4---Precise parsing: the extracted grammar is serialized to some
		parseable form. We use the XML-based interchange format called BGF, or
		BNF-like Grammar Format.
\end{itemize}

The extraction phases are discussed in detail in the following subsections.


\emphsub{Extraction phase 1---Preprocessing}

The first extraction phase, which we call a preprocessing phase, has
the following I/O behavior:

\begin{itemize}
	\item Input: the \xml{<pre>...</pre>} blocks.
	\item Output: a dictionary
		\begin{itemize}
			\item Keys: Left-hand side nonterminals
			\item Values: Arrays of top-level alternatives
		\end{itemize}
\end{itemize}

The phase is subject to the following requirements:

\begin{description}
\item[\textbf{Tag elimination.}] The input notation interleaves tags with proper
  grammar structure. In order to prepare for classic parsing, we need
  to eliminate the tags in the process of constructing properly typed
  lexemes for terminals and nonterminals.
\item[\textbf{Indentation elimination.}] The input notation relies on
  indentation to express top-level choices and line continuation. The
  output format stores top-level choices in arrays, and fuses
  multi-line alternatives.
\item[\textbf{Robustness.}] The inner structure of top-level alternatives is
  parsed simply as a sequence of tokens in the interest of robustness
  so that recovery rules can be applied separately, before, finally,
  the precise grammar structure is parsed.
\end{description}

The preprocessor relies on a stateful scanner (to meet ``tag
elimination'') and a robust parser (to meet ``robustness'').  The
parser recognizes sequences of productions, each one essentially
consisting of a sequence of alternatives; it parses alternatives
as sequences of tokens terminated by CR. The scanner uses three
states:

\begin{itemize}
\item \textbf{italic} upon opening \xml{<i>} tag (or \xml{<em>})
\item \textbf{fixed} upon opening \xml{<code>} tag
\item \textbf{default} when no tag is open
\end{itemize}

That is, we treat each tag as a special token that changes the global
state of the scanner, which in turn can be observed when creating
morphemes for terminals and nonterminals. We also deal with violations
of XML and HTML well-formedness in this manner. The decision
table of the scanner is presented in \autoref{F:decisionTable}
along with  the number of times each decision is taken for
all JLS documents.

\begin{table}
\begin{center}
\small
\begin{tabular}{l|c|c|c|}
&\textit{italic}&\texttt{fixed}&default\\\hline
\tokenAlNum&N (2341
) &T (173
)&T? (194
)\\
\tokenBar&M (2
) &T (2
)&M? (29
)\\
\tokenMeta&M (708
) &T (174
)&T? (200
)\\
\tokenOther&T (198
) &T (165
)&T (205
)\\
\hline\end{tabular}

(T --- terminal, N --- nonterminal, M --- metasymbol)
\end{center}
\label{F:decisionTable}
\caption[Decision table of the extractor's scanner]{Decision table of the extractor's scanner.
Classes of strings are rows, scanner states are columns.}
\end{table}

Most of these decisions are inevitable, even though some of them
pinpoint markup errors. An example of an ``error-free'' decision is to
map an alphanumeric string in the italic mode to a nonterminal. An example
of an ``error-recovering'' decision is to map a non-alphanumeric token
(that does not match any metasymbol) to a terminal---even when it is
tagged with \xml{<i>...</i>}. Several decisions in the ``default''
column involve an element of choice (as indicated by ``?''). The shown
decisions give the best results, that is, they require the least
subsequent transformations of the extracted grammar. For instance, it
turned out that bars without markup were supposed to be BNF bars, but
other metasymbols were better mapped to terminals, whenever markup was
missing. Also, alphanumeric strings without markup turned out to be
mostly terminals, and hence that preference was implemented as a
decision by the scanner.


\emphsub{Extraction phase 2---Error recovery}

We face a few syntax errors with regard to the syntax of the
grammar notation. We also face a number of ``obvious'' semantic
errors in the sense of the language generated by the grammar. We
call them obvious errors because they can be spotted by simple,
generic grammar analyses that involve only very little Java knowledge,
if any. We have opted for an error-recovery approach that relies on a
uniform, rule-based mechanism that performs transformations on each
sequence of tokens that corresponds to an alternative. The rules are
applied until they are no longer applicable. We describe the rules
informally; they are implemented in Python by regular expression matching.

\begin{crule}[Match up parentheses]
  When there is a group (a bar-based choice) that misses an opening or
  closing parenthesis, such as in ``\texttt{(a|b}'', then a nearby
  terminal "(" or ")" (if available) is to be converted to the
  parenthesis, as in \exampleref{JLS2BGF:brackets1}. If there is still
  a closing parenthesis that cannot be matched, then it is dropped, as
  in \exampleref{JLS2BGF:brackets2}.  We have not seen the case of an
  opening parenthesis to remain unmatched, but the rule is implemented
  symmetrically for the sake of completeness.
\end{crule}

\medskip

\begin{example}
\label{JLS2BGF:brackets1}
A grammar production from \citet[\S18.1]{JLS3}: the symbols for closing bracket and parenthesis need to be
  converted to metasymbols to match the opening bracket and parenthesis:
\begin{lstlisting}[language=pp]
TypeArgument:
        Type
        "?" [ ( "extends" | "super" ")" "Type" "]"
\end{lstlisting}
\end{example}

\medskip

\begin{example}
\label{JLS2BGF:brackets2}
A grammar production from \citet[\S18.1]{JLS2} and \citet[\S18.1]{JLS3}: a non-matching square bracket has to be removed:
\begin{lstlisting}[language=pp]
Expression:
        Expression1 [ AssignmentOperator Expression1 ] ]
\end{lstlisting}
\end{example}

\medskip

\begin{crule}[Metasymbol to terminal]
\label{R:m2t}
  (a) When ``$|$'' was scanned as a BNF metasymbol, but it is not used
  in the context of a group, then it is converted to a terminal, as in \exampleref{JLS2BGF:m2t1}.
  
  (b) When ``['' and ``]'' occur next to each other as BNF symbols, then
  they are converted to terminals, as in \exampleref{JLS2BGF:m2t2}.
  
  (c) When ``\{'' and ``\}'' occur
  next to each other as BNF symbols, then they are converted to
  terminals. (Not encountered so far, implemented for the sake of consistency).
  
  (d) When an alternative makes use of the metasymbols for
  grouping, but there is no occurrence of the metasymbol ``$|$'', then
  the parentheses are converted to terminals, as in \exampleref{JLS2BGF:m2t4}.
\end{crule}

\medskip

\begin{example}
\label{JLS2BGF:m2t1}
A grammar production from \citet[\S15.22]{JLS2}: there is no group, so the bar here is not a metasymbol, but a terminal:
\begin{lstlisting}[language=pp]
InclusiveOrExpression:
        ExclusiveOrExpression
        InclusiveOrExpression | ExclusiveOrExpression
\end{lstlisting}
\end{example}

\medskip

\begin{example}
\label{JLS2BGF:m2t2}
A grammar production from \citet[\S8.3]{JLS2}: there is nothing to be made optional, so the square brackets here are not metasymbols, but terminals:
\begin{lstlisting}[language=pp]
VariableDeclaratorId:
        Identifier
        VariableDeclaratorId [ ]
\end{lstlisting}
\end{example}

\medskip

\begin{example}
\label{JLS2BGF:m2t4}
A grammar production from \citet[\S14.19]{JLS2} and \citet[\S18.1]{JLS3}: there is no choice inside the group so the parentheses here are not metasymbols, but terminals:
\begin{lstlisting}[language=pp]
CatchClause:
        "catch" ( FormalParameter ) Block
\end{lstlisting}
\end{example}

\medskip

\begin{crule}[Compose sibling symbols]
\label{R:compose}
  When two alphanumeric nonterminal or terminal tokens are next to
  each other where one of the symbols is of length 1, then they are
  composed as one symbol, as in \exampleref{JLS2BGF:sibling1} and \exampleref{JLS2BGF:sibling2}.
\end{crule}

\medskip

\begin{example}
\label{JLS2BGF:sibling1}
Multiple terminals to compose~\citep[\S19.11]{JLS1}:
\begin{lstlisting}[language=tt]
<code>continu</code><i>e
\end{lstlisting}
\end{example}

\medskip

\begin{example}
\label{JLS2BGF:sibling2}
Multiple nonterminals to compose~\citep[\S14.9]{JLS1}:
\begin{lstlisting}[language=tt]
S<i>witchBlockStatementGroups</i>
\end{lstlisting}
\end{example}

\medskip

\begin{crule}[Decompose compound terminals]
\label{R:decompose}
When a terminal consists of an alphanumeric prefix, followed by ``.'',
possibly followed by a postfix, then the terminal is taken apart
into several ones, as in \exampleref{X:decompose}.
\end{crule}

\medskip

\begin{example}
\label{X:decompose}
Consider this phrase~\citep[\S15.9]{JLS2}:

\begin{lstlisting}[language=tt]
Primary.new Identifier ( ArgumentListopt ) ClassBodyopt
\end{lstlisting}

The decomposition results in the following:

\begin{lstlisting}[language=tt]
Primary . new Identifier ( ArgumentListopt ) ClassBodyopt
\end{lstlisting}
\end{example}

\medskip

\begin{crule}[Nonterminal to terminal]
  Lower-case nonterminals that are not defined by the grammar (i.e., that do
  not occur as a key in the dictionary produced during extraction phase 1), and
  are in lower case, are converted to terminals, as in
  \exampleref{JLS2BGF:nt2t}.
\end{crule}

\medskip

\begin{example}
\label{JLS2BGF:nt2t}
A grammar production from \citet[\S14.11]{JLS2}: \texttt{default} needs to be converted to a terminal:

\begin{lstlisting}[language=tt]
SwitchLabel:
        </em>case<em> ConstantExpression :
        default :
\end{lstlisting}

The same error is present in the later version of the specification \citep[\S14.11]{JLS3}:

\begin{lstlisting}[language=tt]
SwitchLabel:</em>
        case <em>ConstantExpression </em>:
        case <em>EnumConstantName </em>:<em>
        default :
\end{lstlisting}

Note the changes in JLS3: a new alternative was added and the colon was correctly marked up as
a terminal symbol. However, ``\texttt{default}'' is still incorrectly marked up as a nonterminal.
\end{example}

\medskip

\begin{crule}[Terminal to nonterminal]
  Alphanumeric terminals that start in upper case, and are defined by
  the grammar (when considered as nonterminals) are converted, as in \exampleref{JLS2BGF:t2nt}.
\end{crule}

\medskip

\begin{example}
\label{JLS2BGF:t2nt}
A grammar production from \citet[\S7.5]{JLS2}:
\begin{lstlisting}[language=tt]
<em>ImportDeclaration</em>:
        SingleTypeImportDeclaration
        TypeImportOnDemandDeclaration
\end{lstlisting}

The decisive definitions are found in \citet[\S7.5.1, \S7.5.2]{JLS2}:

\medskip

\noindent
\begin{boxedminipage}{0.6\textwidth}
\begin{lstlisting}[language=pp]
SingleTypeImportDeclaration:
        "import" TypeName ";"
TypeImportOnDemandDeclaration:
        "import" PackageOrTypeName "." "*" ";"
\end{lstlisting}
\end{boxedminipage}
\end{example}

\medskip

\begin{crule}[Recover optionality]
  When a nonterminal's name ends on ``opt'', as in ``fooopt'', and the
  grammar defines a nonterminal ``foo'', then the nonterminal
  ``fooopt'' is replaced by $[$foo$]$. (Hence, markup for the subscript
  ``opt'' was missing.)
\end{crule}

\medskip

\begin{example}
Consider again the result of \exampleref{X:decompose}:

\begin{lstlisting}[language=tt]
Primary . new Identifier ( ArgumentListopt ) ClassBodyopt
\end{lstlisting}

After recovery it will be parsed as:

\medskip

\noindent
\begin{boxedminipage}{0.75\textwidth}
\begin{lstlisting}[language=pp]
ClassInstanceCreationExpression:
        Primary "." "new" Identifier "(" ArgumentList? ")" ClassBody?
\end{lstlisting}
\end{boxedminipage}
\end{example}

\medskip

These are all the rules that have stabilized over the project's
duration. Several other rules where investigated but eventually
abandoned because the corresponding issues could be efficiently
addressed by grammar transformations. We used experimental rules to
test for the recurrence of any issue we had spotted. We quantify the
use of the rules shortly.

\bigskip

\emphsub{Extraction phase 3---Removal of doubles}

The JLS documents (deliberately) repeat grammar parts. Hence, we have added a
trivial extraction phase for removal of double alternatives. That is, when a
given right-hand side nonterminal is encountered several times in a source,
then extraction phase 1 accumulates all the alternatives via one entry of the
dictionary, and extraction phase 3 compares alternatives (i.e., sequences of
tokens) to remove any doubles.

\medskip

\begin{example}
\label{JLS2BGF:doubles}
Recall the following definition from \exampleref{JLS2BGF:m2t2}~\citep[\S8.3]{JLS2}:

\begin{lstlisting}[language=tt]
VariableDeclaratorId:
        Identifier
        VariableDeclaratorId [ ]
\end{lstlisting}

The same definition appears elsewhere in the document, even though the markup
is different, but these differences were already neutralized during extraction
phase 1~\citep[\S14.4]{JLS2}:

\begin{lstlisting}[language=tt]
<em>VariableDeclaratorId:
        Identifier
        VariableDeclaratorId</em> [ ]
\end{lstlisting}

Extraction phase 3 preserves 2 alternatives out of 4. As an aside, this
particular example also required the application of \ruleref{R:m2t}.b
because \lstinline[language=tt]{[ ]} must be converted to terminals.
\end{example}


\emphsub{Extraction phase 4---Precise parsing}

Finally, the dictionary structure of extraction phase 1, after the recovery of
extraction phase 2, and double removal of extraction phase 3, is trivially
parsed according to the (E)BNF for the grammar notation, as presented on
\autoref{F:grammar}. In fact, our implementation dumps the extracted grammar
immediately in an XML-based grammar interchange format so that generic grammar
tools for comparison and transformation can take over~\citep{LaemmelZ08}.


\begin{table}[t!]
\begin{center}\normalsize
\begin{tabular}{l|c|c|c|c}
&\numberOfProductions
&\numberOfNonterminals
&\numberOfTops
&\numberOfBottoms
\\\hline\hline

\emph{impl1}&282&135&1&7\\\hline
\emph{read1}&315&148&1&9\\\hline
\emph{impl2}&185&80&6&11\\\hline
\emph{read2}&346&151&1&11\\\hline
\emph{impl3}&245&114&2&12\\\hline
\emph{read3}&435&197&3&14\\\hline
\end{tabular}

\end{center}
\caption{Basic metrics of the JLS grammars.}
\label{F:persource}
\end{table}


\emphsec{JLS grammar metrics}
\label{S:jls-metrics}

\autoref{F:persource} displays simple grammar metrics for the extracted JLS
grammars. A \emph{top nonterminal} is a nonterminal that is defined but never
used; a \emph{bottom nonterminal} is a nonterminal that is used but never
defined~\citep{LaemmelV01b,SellinkV00}. Through continued domain analysis, we
have understood that the major differences between the numbers of productions
and nonterminals for the two grammars of any given version are mainly implied by
the different grammar classes and iteration styles. The decrease of numbers for
the step from \emph{impl1} to \emph{impl2} is explainable with the fact that an
LALR(1) grammar was replaced by a new development (which does not aim at
LALR(1)). Otherwise, the trend is that the numbers of productions and
nonterminals go up with the version number.

The \emph{difference in numbers of top nonterminals is a problem indicator}.
There should be only one top nonterminal: the actual start symbol of the Java
grammar. The \emph{difference in numbers of bottom nonterminals could be
reasonable} because a bottom nonterminal may be a lexeme class---those classes
are somewhat of a grammar design issue. However, a review of the nonterminal
symbols rapidly reveals that some of them correspond to (undefined) categories
of compound syntactic structures.


\begin{table}[t]
\begin{center}
{\small
\begin{tabular}{l|c|c|c|c|c|c||c}

&\textbf{impl1}
&\textbf{impl2}
&\textbf{impl3}
&\textbf{read1}
&\textbf{read2}
&\textbf{read3}
&\textbf{Total}\\\hline\hline
Arbitrary lexical decisions
&
2
&
109
&
60
&
1
&
90
&
161
&
423
\\
\hline
Well-formedness violations
&
5
&
0
&
7
&
4
&
11
&
4
&
31
\\
\hline
Indentation violations
&
1
&
2
&
7
&
1
&
4
&
8
&
23
\\
\hline
Recovery rules
&
3
&
12
&
18
&
2
&
59
&
47
&
141
\\
\myindent Match parentheses
&
0
&
3
&
6
&
0
&
0
&
0
&
9
\\
\myindent Metasymbol to terminal
&
0
&
1
&
7
&
0
&
27
&
7
&
42
\\
\myindent Merge adjacent symbols
&
1
&
0
&
0
&
1
&
1
&
0
&
3
\\
\myindent Split compound symbol
&
0
&
1
&
1
&
0
&
3
&
8
&
13
\\
\myindent Nonterminal to terminal
&
0
&
7
&
3
&
0
&
8
&
11
&
29
\\
\myindent Terminal to nonterminal
&
1
&
0
&
1
&
1
&
17
&
13
&
33
\\
\myindent Recover optionality
&
1
&
0
&
0
&
0
&
3
&
8
&
12
\\
\hline
Purge duplicate definitions
&
0
&
0
&
0
&
16
&
17
&
18
&
51
\\
\hline
Total
&
11
&
123
&
92
&
24
&
181
&
238
&
669
\\
\end{tabular}

}
\end{center}
\caption{Irregularities resolved by grammar extraction.}
\label{F:javatable}
\end{table}


\emphsec{JLS extractor statistics}

Consider \autoref{F:javatable} as an attempt to measure either the
effort needed to complete extraction (manually) or the degree of
inconsistency of the input format. The table summarizes the
frequencies of using recovery rules, handling ``unusual'' continuation
lines\footnote{Our initial guess was that ``substantially indenting''
  means more spaces or tabs than the previous line, but some cases
  were discovered when continuation lines were not indented at all.},
and removal of doubles. The extractor has fixed 669 problems that
otherwise would have prevented straightforward parsing to succeed with
extraction, or implied loss of information, or triggered substantial
grammar transformations.

\section{Grammar transformation}
\label{S:transformation}

In this section we provide an overview of the major
language-independent operators that are needed for the transformation
of concrete syntax definitions in the context of grammar convergence,
we illustrate intended and accidental differences between the JLS
grammars and their representation as operational grammar
transformations, and we summarize the application of our operator
suite to the present JLS study.

\emphsec{Operator suite}
\label{S:xbgf}

\begin{figure}[t!]
\begin{boxedminipage}{\hsize}
\begin{center}
\includegraphics[width=.98\textwidth]{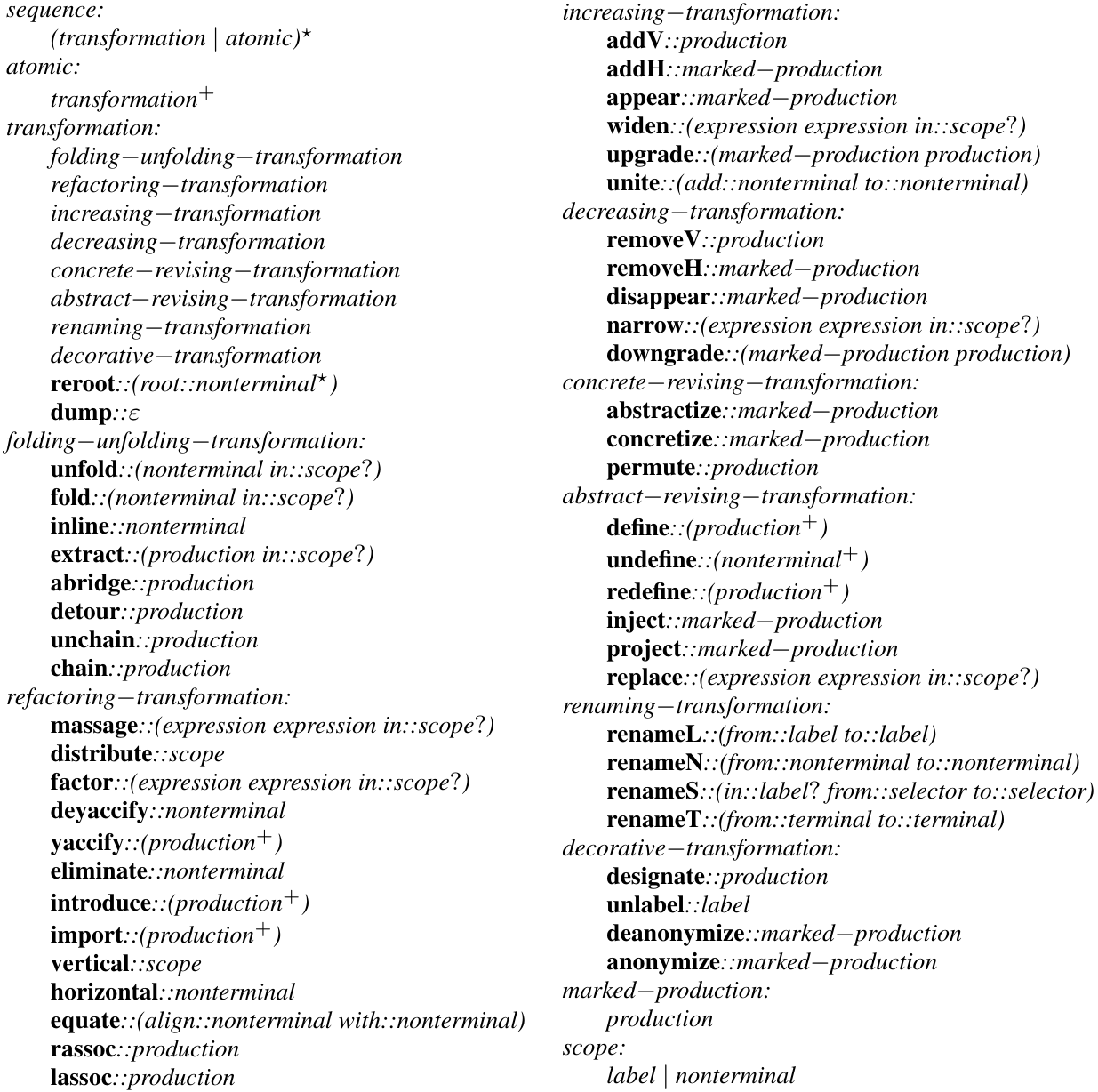}
\end{center}
\end{boxedminipage}
  \caption{The grammar of the transformation language.
Metasyntax nonterminals like \emph{production} and \emph{expression} are deliberately omitted for clarity.}
    \label{F:xbgf}
\end{figure}

Just as in \citet{LaemmelZ08}, we distinguish here among
semantics-preserving, semantics-increasing, semantics-decreasing and
semantics-editing operators. The term semantics refers to the language
generated by the grammar---when considered as a set of strings.

The complete grammar of our grammar transformation language is presented on
\autoref{F:xbgf}. As we see, many operators have the form of either $f(n)$,
where $f$ is the operator in question, and $n$ is the nonterminal to be
affected, or $f(x,y)$, where $f$ is the operator in question, $x$ is the
grammar expression to be located in the input, and $y$ is the corresponding
replacement. There are also several operators that are parametrized with a so
called ``marked production''---a grammar production that has parts of it
specifically marked as local transformation targets.


Folding, unfolding and refactoring operators preserve the string-oriented semantics. Many
of them are implemented as conditional replacements. For instance,
\textbf{massage} has a pre-condition that the two expressions that
parametrize it are $massage$-equal. If the pre-condition is satisfied, the former
expression is replaced with the latter; if it fails, the
transformation chain stops with an error reported. The
$massage$-equality is defined by a set of algebraic laws listed in
\autoref{A:massage} for completeness' sake.


Below we briefly explain the operators that are needed to understand the present JLS study; see also \autoref{F:xbgfpercommand}.

\begin{itemize}
	\item \textbf{unfold} replaces all occurrences of a given nonterminal by its definition,
		possibly limited by scope.
	\item \textbf{fold} replaces all occurrences of a given nonterminal's definition by the nonterminal itself,
		possibly limited by scope.
	\item \textbf{inline} is the same as \textbf{unfold}, but it also removes the nonterminal from the grammar
		after unfolding (if some recursion prevents it, the operator fails).
	\item \textbf{extract} introduces a new nonterminal to the grammar and \textbf{fold}s its definition.
	\item \textbf{chain} introduces a chain production by performing a local \textbf{extract} on the
		whole definition of a given nonterminal (i.e., it goes from $a:xyz$ to $a:b$ and $b:xyz$).
	\item \textbf{massage} allows for rewriting grammar expressions according to the algebraic laws listed in \autoref{A:massage}.
	\item \textbf{distribute} pulls the choices that occur inside a grammar expression outwards.
	\item \textbf{factor} rewrites an expression to an equivalent but differently factored one.
		Its common use is to push choices deeper since an automated \textbf{distribute} operator is more useful in other cases.
	\item \textbf{deyaccify} converts a recursive definition-based style of iteration to the use
	 	of the EBNF operators ``+'' and ``*''. The operator name is justified by \citet{JongeM01} and \citet{Laemmel01a}.
	\item \textbf{yaccify} is the intended inverse of \textbf{deyaccify}.
	\item \textbf{eliminate} removes a production from the grammar if it is not used anywhere, otherwise fails.
	\item \textbf{introduce} adds a free nonterminal with its definition to the grammar.
	\item \textbf{import} works as multiple \textbf{introduce} operators for introducing possibly interconnected productions.
	\item \textbf{vertical} converts a ``horizontal'' production with top-level choices
		(previously called ``flat'' by \citet{LaemmelW01}) to several productions.
	\item \textbf{horizontal} converts a ``vertical'' nonterminal definition consisting of multiple productions
			(previously called ``non-flat'' by \citet{LaemmelW01}) to a single production
			with top-level choices.
	\item \textbf{addV} adds a production to the existing nonterminal definition.
	\item \textbf{addH} replaces any expression with a choice involving that expression and something else
		(i.e., it goes from $a:xyz$ to $a:x(y|b)z$).
	\item \textbf{appear} injects a nillable symbol (the one that can be evaluated to $\varepsilon$).
	\item \textbf{widen} replaces an expression with a more general one
		(e.g., $x^+$ with $x^\star$).
	\item \textbf{upgrade} replaces an expression with a nonterminal that can possibly be evaluated to it.
	\item \textbf{unite} merges the definitions and the uses of two nonterminals.
	\item \textbf{removeV} removes a production from the existing nonterminal definition such that
		the nonterminal does not become undefined.
	\item \textbf{removeH} replaces any choice with a similar choice with less alternatives.
	\item \textbf{disappear} projects a nillable symbol (the one that can be evaluated to $\varepsilon$).
	\item \textbf{narrow} replaces an expression with a refined one.
	\item \textbf{downgrade} replaces a nonterminal with one of its definitions.
	\item \textbf{define} adds a definition to a bottom (used but undefined) nonterminal.
	\item \textbf{undefine} removes a definition of a nonterminal, forcing it to become a bottom sort.
	\item \textbf{redefine} replaces the existing definition of a nonterminal with the given one.
	\item \textbf{inject} inserts possibly non-nillable symbols to a sequence.
	\item \textbf{project} removes possibly non-nillable symbols from a sequence.
	\item \textbf{replace} replaces any subexpression with another one.
	\item \textbf{unlabel} strips a production of its label.
\end{itemize}

\emphsec{Examples of grammar transformations}


The examples that illustrate the uses of the XBGF operators are presented below in
the same order they are encountered in the transformation scripts with respect to
the convergence phases that we listed and motivated in \S\ref{S:process}.

\emphsubsub{Convergence preparation}
\label{X:preparation}

\noskip The following two examples pinpoint ``grammar bugs'': incorrect syntax. In
some cases, incorrect syntax merely arises from representation anomalies of the
HTML input used for extraction---as shown below.

\noindent\begin{minipage}{\textwidth}
\begin{flushright}
	\emph{impl3} \citep[\S18.1]{JLS3}
\end{flushright}
\noskip\noindent
\begin{boxedminipage}{\textwidth}
\begin{lstlisting}[language=pp]
Block:
        BlockStatements*
\end{lstlisting}
\end{boxedminipage}
\end{minipage}

\noindent\begin{minipage}{\textwidth}
\begin{flushright}
	Semantics-revising transformation
\end{flushright}
\noskip\noindent
\begin{boxedminipage}{\textwidth}
\begin{lstlisting}[language=pp]
replace(
  BlockStatements* ,
  "{" BlockStatements "}" );
\end{lstlisting}
\end{boxedminipage}
\end{minipage}

\noindent\begin{minipage}{\textwidth}
\begin{flushright}
	Corrected syntax
\end{flushright}
\noskip\noindent
\begin{boxedminipage}{\textwidth}
\begin{lstlisting}[language=pp]
Block:
        "{" BlockStatements "}"
\end{lstlisting}
\end{boxedminipage}
\end{minipage}

The source format defines curly brackets to express iteration. However, in the
example at hand, taken from \emph{impl3}, they were meant as terminal symbols, and
were not recognized due to missing markup. The incorrect list construct is replaced
accordingly.



Another activity frequently undertaken during the convergence preparation phase is
initial correction. Unlike correction that happens during convergence resolution
phase (see \S\ref{X:correction}), it is triggered by an external bug report and not
by a grammar comparator. Such a bug report can be produced by another tool or taken
from a third party. For instance, a misnamed nonterminal is found in \emph{impl2}
when examining the list of bottom nonterminals before the later phases of
convergence process start:

\noindent\begin{minipage}{\textwidth}
\begin{flushright}
	\emph{impl2} \citep[\S18.1]{JLS2}
\end{flushright}
\noskip\noindent
\begin{boxedminipage}{\textwidth}
\begin{lstlisting}[language=pp]
Expression3:
        "(" (Expr | Type) ")" Expression3
\end{lstlisting}
\end{boxedminipage}
\end{minipage}

\noindent\begin{minipage}{\textwidth}
\begin{flushright}
	Semantics-revising transformation
\end{flushright}
\noskip\noindent
\begin{boxedminipage}{\textwidth}
\begin{lstlisting}[language=pp]
replace(
	 Expr,
	 Expression);
\end{lstlisting}
\end{boxedminipage}
\end{minipage}

\noindent\begin{minipage}{\textwidth}
\begin{flushright}
	Corrected syntax
\end{flushright}
\noskip\noindent
\begin{boxedminipage}{\textwidth}
\begin{lstlisting}[language=pp]
Expression3:
        "(" (Expression | Type) ")" Expression3
\end{lstlisting}
\end{boxedminipage}
\end{minipage}

\emphsubsub{Nominal matching}
\label{X:nominal-matching}


\noskip
In \emph{impl2}, the built-in variable types are defined by a nonterminal
called \emph{BasicType} which contains all the alternatives. In the more
readable counterpart the same nonterminal is called \emph{PrimitiveType},
and it requires several intermediate nonterminals because types are explained
in two different language document sections.

\medskip

\noindent\begin{minipage}{\textwidth}
\begin{flushright}
	\emph{read2} \citep[\S4.1, \S4.2]{JLS2}
\end{flushright}
\noskip\noindent
\begin{boxedminipage}{\textwidth}
\begin{lstlisting}[language=pp]
PrimitiveType:
        NumericType 
        "boolean" 
NumericType:
        IntegralType 
        FloatingPointType 
IntegralType:
        "byte" 
        "short" 
        "int" 
        "long" 
        "char" 
FloatingPointType:
        "float" 
        "double" 
\end{lstlisting}
\end{boxedminipage}
\end{minipage}

\noindent\begin{minipage}{\textwidth}
\begin{flushright}
	\emph{impl2} \citep[\S18.1]{JLS2}
\end{flushright}
\noskip\noindent
\begin{boxedminipage}{\textwidth}
\begin{lstlisting}[language=pp]
BasicType:
        "byte" 
        "short" 
        "char" 
        "int" 
        "long" 
        "float" 
        "double" 
        "boolean" 
\end{lstlisting}
\end{boxedminipage}
\end{minipage}

\noindent\begin{minipage}{\textwidth}
\begin{flushright}
	Semantics-preserving transformation
\end{flushright}
\noskip\noindent
\begin{boxedminipage}{\textwidth}
\begin{lstlisting}[language=pp]
renameN(PrimitiveType, BasicType);
inline(IntegralType);
inline(FloatingPointType);
inline(NumericType);
distribute(BasicType);
\end{lstlisting}
\end{boxedminipage}
\end{minipage}

\medskip

The last \textbf{distribute} transformation is needed to normalize the definition
of \emph{BasicType} to a top-level choice.

\emphsubsub{Structural matching}
\label{X:matching}


\noskip In \emph{read2}, there are distinct alternatives for blocks vs.\ static
blocks. In contrast, in \emph{impl2}, these forms appear in a factored manner.
Hence, the \textbf{factor} operator is used in the following example to factor out
the shared reference to \emph{Block}. Then, the \textbf{massage} operator changes
the style of expressing optionality of the keyword ``\texttt{static}''.

\noindent\begin{minipage}{\textwidth}
\begin{flushright}
	\emph{read2} \citep[\S8.1.5, \S8.6, \S8.7]{JLS2}
\end{flushright}
\noskip\noindent
\begin{boxedminipage}{\textwidth}
\begin{lstlisting}[language=pp]
ClassBodyDeclaration:
        InstanceInitializer
        StaticInitializer
        ...
StaticInitializer:
        "static" Block
InstanceInitializer:
        Block
\end{lstlisting}
\end{boxedminipage}
\end{minipage}

\noindent\begin{minipage}{\textwidth}
\begin{flushright}
	\emph{impl2} \citep[\S18.1]{JLS2}
\end{flushright}
\noskip\noindent
\begin{boxedminipage}{\textwidth}
\begin{lstlisting}[language=pp]
ClassBodyDeclaration:
        "static"? Block
\end{lstlisting}
\end{boxedminipage}
\end{minipage}

\noindent\begin{minipage}{\textwidth}
\begin{flushright}
	Semantics-preserving transformation
\end{flushright}
\noskip\noindent
\begin{boxedminipage}{\textwidth}
\begin{lstlisting}[language=pp]
inline(StaticInitializer);
inline(InstanceInitializer);
factor(
  (Block | ("static" Block)) ,
  ((EPSILON | "static") Block) );
massage(
  (EPSILON | "static") ,
  "static"? );
\end{lstlisting}
\end{boxedminipage}
\end{minipage}

\medskip



Like the previous transformation sample, the following one is taken from a
refactoring script that aligns \emph{read2} with \emph{impl2}. The JLS case
involves many hundreds of such small refactoring steps; see \S\ref{S:measures}.

In \emph{read2}, the recursion-based style of iteration is used. For instance,
there is a recursively defined nonterminal \emph{ClassBodyDeclarations} for lists
of \emph{ClassBodyDeclaration}. In contrast, in \emph{impl2}, the list form ``*''
is used. Deyaccification replaces the recursive definition of
\emph{ClassBodyDeclarations} by \emph{ClassBodyDeclaration}$^+$. The nonterminal
\emph{ClassBodyDeclarations} is no longer needed, and hence inlined. The list of
declarations was optional, and hence ``+'' and ``?'' can be simplified to ``*''.

\noindent\begin{minipage}{\textwidth}
\begin{flushright}
	\emph{read2} \citep[\S8.1.5]{JLS2}
\end{flushright}
\noskip\noindent
\begin{boxedminipage}{\textwidth}
\begin{lstlisting}[language=pp]
ClassBody:
        "{" ClassBodyDeclarations? "}"
ClassBodyDeclarations:
        ClassBodyDeclaration
        ClassBodyDeclarations ClassBodyDeclaration
\end{lstlisting}
\end{boxedminipage}
\end{minipage}

\noindent\begin{minipage}{\textwidth}
\begin{flushright}
	\emph{impl2} \citep[\S18.1]{JLS2}
\end{flushright}
\noskip\noindent
\begin{boxedminipage}{\textwidth}
\begin{lstlisting}[language=pp]
ClassBody:
        "{" ClassBodyDeclaration* "}"
\end{lstlisting}
\end{boxedminipage}
\end{minipage}

\noindent\begin{minipage}{\textwidth}
\begin{flushright}
	Semantics-preserving transformation
\end{flushright}
\noskip\noindent
\begin{boxedminipage}{\textwidth}
\begin{lstlisting}[language=pp]
deyaccify(ClassBodyDeclarations);
inline(ClassBodyDeclarations);
massage(
  ClassBodyDeclaration+? ,
  ClassBodyDeclaration* );
\end{lstlisting}
\end{boxedminipage}
\end{minipage}


\medskip

\emphsubsub{Extension}
\label{X:extension}


\noskip The following transformation is part of a chain that captures the difference
between JLS1 and JLS2. The particular widening step enables \emph{instance}
initializers in class bodies where only \emph{static} initializers were admitted
before.

The example also demonstrates that transformation operators may carry an extra
argument to describe the \emph{scope of replacement} (recall \autoref{F:xbgf}). By
default, the scope is universal: all matching expressions in the input grammar
would be affected. Selective scopes are nonterminal definitions (specified by a
nonterminal---as in the following example) or productions (specified by a
production label).

\medskip

\noindent\begin{minipage}{\textwidth}
\begin{flushright}
	\emph{jls1} after many transformation steps \citep{JLS1}
\end{flushright}
\noskip\noindent
\begin{boxedminipage}{\textwidth}
\begin{lstlisting}[language=pp]
ClassBodyDeclaration:
        "static" Block
\end{lstlisting}
\end{boxedminipage}
\end{minipage}

\noindent\begin{minipage}{\textwidth}
\begin{flushright}
	\emph{impl2} \citep[\S9.3]{JLS2}
\end{flushright}
\noskip\noindent
\begin{boxedminipage}{\textwidth}
\begin{lstlisting}[language=pp]
ClassBodyDeclaration:
        "static"? Block
\end{lstlisting}
\end{boxedminipage}
\end{minipage}

\noindent\begin{minipage}{\textwidth}
\begin{flushright}
	Semantics-increasing transformation
\end{flushright}
\noskip\noindent
\begin{boxedminipage}{\textwidth}
\begin{lstlisting}[language=pp]
widen(
  "static",
  "static"?,
  in ClassBodyDeclaration);
\end{lstlisting}
\end{boxedminipage}
\end{minipage}


The following transformation is part of a script that captures the difference between
JLS2 and JLS3, where the latter offers \emph{Annotation} as the additional option.

\noindent\begin{minipage}{\textwidth}
\begin{flushright}
	\emph{read2} \citep[\S9.3]{JLS2}
\end{flushright}
\noskip\noindent
\begin{boxedminipage}{\textwidth}
\begin{lstlisting}[language=pp]
ConstantModifier:
        "public"
        "static"
        "final"
\end{lstlisting}
\end{boxedminipage}
\end{minipage}

\noindent\begin{minipage}{\textwidth}
\begin{flushright}
	\emph{read3} \citep[\S9.3]{JLS3}
\end{flushright}
\noskip\noindent
\begin{boxedminipage}{\textwidth}
\begin{lstlisting}[language=pp]
ConstantModifier:
        Annotation
        "public"
        "static"
        "final"
\end{lstlisting}
\end{boxedminipage}
\end{minipage}

\noindent\begin{minipage}{\textwidth}
\begin{flushright}
	Semantics-increasing transformation
\end{flushright}
\noskip\noindent
\begin{boxedminipage}{\textwidth}
\begin{lstlisting}[language=pp]
addV(
 ConstantModifier:
        Annotation
);
\end{lstlisting}
\end{boxedminipage}
\end{minipage}

When we seek relationships between grammars of different versions, then
semantics-increasing/-decreasing transformations are clearly to be expected. As a
matter of discipline, we prefer to describe the difference by a semantic-increasing
transformation to map a version to its successor version (as opposed to the inverse
direction). We speak of \emph{grammar extension} in this case.

\emphsubsub{Relaxation}
\label{X:relaxation}


Increase (or decrease) may also be needed when two grammars
are essentially equivalent---except that one is more permissive than
the other. This actually happens in practice: a permissive grammar may
be needed as a concession to practicality of, say, parser
implementation. We also speak of \emph{grammar relaxation} in this case. In
the JLS case, the different purposes of the grammars (to be more or
less readable or implementable respectively) imply the need for relaxation.
Similar issues arise with relationships between abstract and concrete
syntaxes~\citep{LaemmelZ08,Wile97b}.

In \emph{impl2}, there is only one category of (arbitrary) modifiers. In contrast,
in \emph{read2}, there are various precise categories of modifiers for classes,
fields, methods, constructors, interfaces, constants and abstract methods.
Accordingly, the \emph{impl2} grammar is more permissive as far as modifiers are
concerned. The grammar fragments in the following example are deliberately
pretty-printed as horizontal productions for the sake of readability. In reality
the extractor produces only vertical ones as usual.

\smallskip

\noindent\begin{minipage}{\textwidth}
\begin{flushright}
	\emph{read2} \citep[\S8.1.1, \S8.3.1, \S8.4.3, \S8.8.3, \S9.1.1, \S9.3, \S9.4]{JLS2}
\end{flushright}
\noskip\noindent
\begin{boxedminipage}{\textwidth}
\begin{lstlisting}[language=pp]
ClassModifier:
        "public" | "protected" | "private" | "abstract" | "static"
                | "final" | "strictfp"
FieldModifier:
        "public" | "protected" | "private" | "static" | "final" |
                "transient" | "volatile"
MethodModifier:
        "public" | "protected" | "private" | "abstract" | "static"
                | "final" | "synchronized" | "native" | "strictfp"
ConstructorModifier:
        "public" | "protected" | "private"
InterfaceModifier:
        "public" | "protected" | "private" | "abstract" | "static"
                | "strictfp"
ConstantModifier:
        "public" | "static" | "final"
AbstractMethodModifier:
        "public" | "abstract"
\end{lstlisting}
\end{boxedminipage}
\end{minipage}

\noindent\begin{minipage}{\textwidth}
\begin{flushright}
	\emph{impl2} \citep[\S18.1]{JLS2}
\end{flushright}
\noskip\noindent
\begin{boxedminipage}{\textwidth}
\begin{lstlisting}[language=pp]
Modifier:
        "public" | "protected" | "private" | "static" | "abstract"
                | "final" | "native" | "synchronized" | "transient"
                | "volatile" | "strictfp"
\end{lstlisting}
\end{boxedminipage}
\end{minipage}

\noindent\begin{minipage}{\textwidth}
\begin{flushright}
	Semantics-increasing transformation
\end{flushright}

\noindent
\begin{boxedminipage}{\textwidth}
\begin{lstlisting}[language=pp]
	renameN(ClassModifier, Modifier);
	unite(FieldModifier, Modifier);
	unite(MethodModifier, Modifier);
	unite(ConstructorModifier, Modifier);
	unite(InterfaceModifier, Modifier);
	unite(ConstantModifier, Modifier);
	unite(AbstractMethodModifier, Modifier);
\end{lstlisting}
\end{boxedminipage}
\end{minipage}

\medskip


Constructor declarations are defined very differently in \emph{impl3} and
\emph{read3}. The following example shows the last steps of their convergence,
where \emph{ConstructorBody} must be replaced by \emph{MethodBody}. However, the
definition of \emph{ConstructorBody} (at this stage) is equal to the definition of
\emph{Block}, while \emph{MethodBody} can be \emph{Block} or \texttt{";"}. By
letting the terminal \texttt{";"} appear instead of \emph{Block}, we make the
grammar of the language more permissive.

\smallskip

\noindent\begin{minipage}{\textwidth}
\begin{flushright}
	\emph{read3} after many transformation steps
\end{flushright}
\noskip\noindent
\begin{boxedminipage}{\textwidth}
\begin{lstlisting}[language=pp]
ConstructorDeclaratorRest:
        FormalParameters Throws? ConstructorBody
ConstructorBody:
        "{" BlockStatements "}"
MethodBody:
        Block
MethodBody:
        ";"
Block:
        "{" BlockStatements "}"
\end{lstlisting}
\end{boxedminipage}
\end{minipage}

\noindent\begin{minipage}{\textwidth}
\begin{flushright}
	\emph{impl3} \citep[\S18.1]{JLS3}
\end{flushright}
\noskip\noindent
\begin{boxedminipage}{\textwidth}
\begin{lstlisting}[language=pp]
ConstructorDeclaratorRest:
        FormalParameters ("throws" QualifiedIdentifierList)? MethodBody
\end{lstlisting}
\end{boxedminipage}
\end{minipage}

\noindent\begin{minipage}{\textwidth}
\begin{flushright}
	Semantics-increasing transformation
\end{flushright}
\noskip\noindent
\begin{boxedminipage}{\textwidth}
\begin{lstlisting}[language=pp]
fold(Block);
upgrade(
 ConstructorBody:
        <MethodBody>
 MethodBody:
        Block
);
inline(ConstructorBody);
inline(Throws);
\end{lstlisting}
\end{boxedminipage}
\end{minipage}

We suggest that a language specification should explicitly call out
relaxations so that they are not confused with overlooked
inconsistencies (to be modeled as corrections) or evolutionary
differences (to be modeled as extensions).


\medskip

\emphsubsub{Correction}
\label{X:correction}


Finally, two grammars may differ (with regard to the generated
language) in a manner that is purely accidental (read as
``incorrect''). We speak of (transformations for) \emph{grammar
  correction} in this case. 

What Oracle SDN (Sun Developer Network) Bug Database
reports\footnote{\url{http://bugs.sun.com/bugdatabase/view_bug.do?bug_id=6442525}}
as ``the master bug for errors in the Java grammar'' is the fact that \S18.1 of
\citet{JLS3} does not permit the obsolescent array syntax in a method declaration
of an annotation type.

\noindent\begin{minipage}{\textwidth}
\begin{flushright}
	Incorrect syntax in \emph{impl3}
\end{flushright}
\noskip\noindent
\begin{boxedminipage}{\textwidth}
\begin{lstlisting}[language=pp]
AnnotationMethodRest:
        "(" ")" DefaultValue?
\end{lstlisting}
\end{boxedminipage}
\end{minipage}

\noindent\begin{minipage}{\textwidth}
\begin{flushright}
	Semantics-increasing transformation
\end{flushright}
\noskip\noindent
\begin{boxedminipage}{\textwidth}
\begin{lstlisting}[language=pp]
appear(
 AnnotationMethodRest:
        "(" ")" <("[" "]")*> DefaultValue?
);
\end{lstlisting}
\end{boxedminipage}
\end{minipage}

\noindent\begin{minipage}{\textwidth}
\begin{flushright}
	Corrected syntax
\end{flushright}
\noskip\noindent
\begin{boxedminipage}{\textwidth}
\begin{lstlisting}[language=pp]
AnnotationMethodRest:
        "(" ")" ("[" "]")* DefaultValue?
\end{lstlisting}
\end{boxedminipage}
\end{minipage}


\medskip


Not all corrections may be expressed in terms
of semantics-increasing/-decreasing operators. If that is not possible, we have to
use less disciplined operators.
For example, the production for the \emph{break} statement in \emph{impl2} lacks the semicolon
which is injected accordingly (left unnoticed in Bosworth's bug list,
but obvious when converging with \emph{read2}).

\noindent\begin{minipage}{\textwidth}
\begin{flushright}
	\emph{impl2} \citep[\S18.1]{JLS2}
\end{flushright}
\noskip\noindent
\begin{boxedminipage}{\textwidth}
\begin{lstlisting}[language=pp]
Statement:
        "break" Identifier?
\end{lstlisting}
\end{boxedminipage}
\end{minipage}

\noindent\begin{minipage}{\textwidth}
\begin{flushright}
	Semantics-revising transformation
\end{flushright}
\noskip\noindent
\begin{boxedminipage}{\textwidth}
\begin{lstlisting}[language=pp]
inject(
 Statement:
        "break" Identifier? < ";" >
);
\end{lstlisting}
\end{boxedminipage}
\end{minipage}

\noindent\begin{minipage}{\textwidth}
\begin{flushright}
	Corrected syntax
\end{flushright}
\noskip\noindent
\begin{boxedminipage}{\textwidth}
\begin{lstlisting}[language=pp]
Statement:
        "break" Identifier? ";"
\end{lstlisting}
\end{boxedminipage}
\end{minipage}


The \emph{impl2} and \emph{impl3} grammars define the Java expression
syntax by means of layers, i.e., there are several nonterminals
  \emph{Expression1}, \emph{Expression2}, ... for the different
  priorities. We are concerned with one layer here. The second rule
  for \emph{Expression2Rest} contains an offending occurrence of
  \emph{Expression3} which needs to be projected away. This issue was
  revealed by comparing the \emph{impl2} and \emph{impl3} grammars
  with the \emph{read2} and \emph{read3} grammars
  (after some prior refactoring).

\noindent\begin{minipage}{\textwidth}
\begin{flushright}
	Incorrect expression syntax in \emph{impl2} and \emph{impl3}
\end{flushright}
\noskip\noindent
\begin{boxedminipage}{\textwidth}
\begin{lstlisting}[language=pp]
Expression2:
        Expression3 [ Expression2Rest ]
Expression2Rest:
        (InfixOp Expression3)*
Expression2Rest:
        Expression3 "instanceof" Type
\end{lstlisting}
\end{boxedminipage}
\end{minipage}

\noindent\begin{minipage}{\textwidth}
\begin{flushright}
	Semantics-revising transformation
\end{flushright}
\noskip\noindent
\begin{boxedminipage}{\textwidth}
\begin{lstlisting}[language=pp]
project(
  Expression2Rest:
    < Expression3 > "instanceof" Type
);
\end{lstlisting}
\end{boxedminipage}
\end{minipage}

\noindent\begin{minipage}{\textwidth}
\begin{flushright}
	Corrected syntax
\end{flushright}
\noskip\noindent
\begin{boxedminipage}{\textwidth}
\begin{lstlisting}[language=pp]
Expression2:
        Expression3 [ Expression2Rest ]
Expression2Rest:
        (InfixOp Expression3)*
Expression2Rest:
        "instanceof" Type
\end{lstlisting}
\end{boxedminipage}
\end{minipage}

\smallskip


\emphsec{Postmortem of the JLS case}
\label{S:postmortem}

\begin{table}[t!]
\begin{center}\small
\begin{tabular}{l|c|c|c|c}
&\numberOfProductions
&\numberOfNonterminals
&\numberOfTops
&\numberOfBottoms
\\\hline\hline

\emph{jls1}&278&132&1&7\\\hline
\emph{jls2}&178&75&1&7\\\hline
\emph{jls3}&236&109&1&7\\\hline
\emph{jls12}&178&75&1&7\\\hline
\emph{jls123}&236&109&1&7\\\hline
\emph{read12}&345&152&1&7\\\hline
\emph{read123}&438&201&1&7\\\hline
\end{tabular}

\end{center}
\caption{Simple metrics for the derived JLS grammars.}
\medskip
\label{F:pertarget}
\end{table}

\begin{table}[t!]
\begin{center}
{\scriptsize
\begin{tabular}{l|c|c|c|c|c|c|c||c}
&\textbf{jls1} &\textbf{jls12} &\textbf{jls123} &\textbf{jls2} &\textbf{jls3} &\textbf{read12} &\textbf{read123} &\textbf{Total}\\\hline
\javaNumberOfLines { 682 }
{ 5114 }
{ 2847 }
{ 6774 }
{ 10721 }
{ 1639 }
{ 3082 }
{30859}
\hline
\javaNumberOfTransformations { 67 }
{ 290 }
{ 111 }
{ 387 }
{ 544 }
{ 77 }
{ 135 }
{1611}
\javaNumberOfRefactors { 45 }
{ 231 }
{ 80 }
{ 275 }
{ 381 }
{ 31 }
{ 78 }
{1121}
\javaNumberOfGeneralises { 22 }
{ 58 }
{ 31 }
{ 102 }
{ 150 }
{ 39 }
{ 53 }
{455}
\javaNumberOfRevisings {---} { 1 }
{---} { 10 }
{ 13 }
{ 7 }
{ 4 }
{35}
\hline
\javaNumberOfSteps { 8 }
{ 9 }
{ 8 }
{ 22 }
{ 26 }
{ 6 }
{ 6 }
{85}
\javaNumberOfIssues { 8 }
{ 25 }
{ 17 }
{ 40 }
{ 53 }
{ 32 }
{ 44 }
{219}
\javaIssuesPostX {---} {---} {---} { 7 }
{ 8 }
{ 7 }
{ 4 }
{26}
\javaIssuesCorrect { 8 }
{ 5 }
{---} { 45 }
{ 53 }
{ 14 }
{ 14 }
{139}
\javaIssuesExtend {---} { 17 }
{ 14 }
{ 1 }
{---} { 14 }
{ 29 }
{75}
\javaIssuesPermit { 3 }
{ 6 }
{ 3 }
{ 9 }
{ 19 }
{ 1 }
{---} {41}
\javaEarly { 1 }
{---} {---} { 15 }
{ 24 }
{ 11 }
{ 14 }
{65}
\javaEarlyKnownBugs {---} {---} {---} { 1 }
{ 11 }
{---} { 4 }
{16}
\javaEarlyPostExtraction {---} {---} {---} { 7 }
{ 8 }
{ 7 }
{ 5 }
{27}
\javaEarlyInitialCorrection { 1 }
{---} {---} { 7 }
{ 5 }
{ 4 }
{ 5 }
{22}
\hline
\javaFinal { 21 }
{ 59 }
{ 31 }
{ 97 }
{ 139 }
{ 35 }
{ 43 }
{425}
\javaFinalExtension {---} { 17 }
{ 26 }
{---} {---} { 31 }
{ 38 }
{112}
\javaFinalRelaxation { 18 }
{ 39 }
{ 5 }
{ 75 }
{ 112 }
{---} { 2 }
{251}
\javaFinalCorrection { 3 }
{ 3 }
{---} { 22 }
{ 27 }
{ 4 }
{ 3 }
{62}
\hline
\end{tabular}

}
\end{center}
\caption{Transformation of the JLS grammars---effort metrics and categorization.}
\medskip
\label{F:xbgfeffort}
\end{table}

\begin{table}
\begin{center}
{\small
\begin{tabular}{l|c|c|c|c|c|c|c||c}
&\textbf{jls1} &\textbf{jls12} &\textbf{jls123} &\textbf{jls2} &\textbf{jls3} &\textbf{read12} &\textbf{read123} &\textbf{Total}\\\hline
\xbgfNumber{rename}&9&4&2&9&10& ---&2&36\\
\xbgfNumber{reroot}&2& ---& ---&2&2&2&1&9\\
\hline
\xbgfNumber{unfold}&1&10&8&11&13&2&3&48\\
\xbgfNumber{fold}&4&11&4&11&13&2&5&50\\
\xbgfNumber{inline}&3&67&8&71&100& ---&1&250\\
\xbgfNumber{extract}& ---&17&5&18&30& ---&5&75\\
\xbgfNumber{chain}&1& ---&2& ---& ---&1&4&8\\
\hline
\xbgfNumber{massage}&2&13& ---&15&32&5&3&70\\
\xbgfNumber{distribute}&3&4&2&3&6& ---& ---&18\\
\xbgfNumber{factor}&1&7&3&5&24&3&1&44\\
\xbgfNumber{deyaccify}&2&20& ---&25&33&4&3&87\\
\xbgfNumber{yaccify}& ---& ---& ---& ---&1& ---&1&2\\
\xbgfNumber{eliminate}&1&8&1&14&22& ---& ---&46\\
\xbgfNumber{introduce}& ---&1&30&4&13&3&34&85\\
\xbgfNumber{import}& ---& ---&2& ---& ---& ---&1&3\\
\xbgfNumber{vertical}&5&7&7&8&22&5&8&62\\
\xbgfNumber{horizontal}&4&19&5&17&31&4&4&84\\
\hline
\xbgfNumber{add}&1&14&13&7&20&28&20&103\\
\xbgfNumber{appear}& ---&8&11&8&25&2&17&71\\
\xbgfNumber{widen}&1&3& ---&1&8&1&3&17\\
\xbgfNumber{upgrade}& ---&8& ---&14&20&2&2&46\\
\xbgfNumber{unite}&18&2& ---&18&21&5&4&68\\
\hline
\xbgfNumber{remove}& ---&10&1&11&18& ---&1&41\\
\xbgfNumber{disappear}& ---&7&4&11&11& ---& ---&33\\
\xbgfNumber{narrow}& ---& ---&1& ---&4& ---& ---&5\\
\xbgfNumber{downgrade}& ---&2& ---&8&3& ---& ---&13\\
\hline
\hline
\xbgfNumber{define}& ---&6& ---&4&9&1&6&26\\
\xbgfNumber{undefine}& ---&3& ---&5&3& ---& ---&11\\
\xbgfNumber{redefine}& ---&3& ---&8&7&6&2&26\\
\xbgfNumber{inject}& ---& ---& ---&2&4& ---&1&7\\
\xbgfNumber{project}& ---&1& ---&1&2& ---& ---&4\\
\xbgfNumber{replace}&3&1&2&3&6&1&1&17\\
\hline
\xbgfNumber{unlabel}& ---& ---& ---& ---& ---& ---&2&2\\
\hline
\end{tabular}

}
\end{center}
\caption{XBGF operators usage for JLS convergence.}
\label{F:xbgfpercommand}
\end{table}

\label{S:measures}

We recall that \autoref{F:persource} lists simple metrics for the leaves of
JLS' convergence tree. The new \autoref{F:pertarget} shows the same data for
the derived grammars. It is easily seen that top- and bottom-nonterminals are
consolidated now. In the case of the ``common denominators'' \emph{jls1--3},
the numbers of nonterminals and productions reflect that these grammars were
derived to be similar to \emph{impl1--3}. Similar correlations hold for the
``inter-version'' grammars in the rest of the table.

\autoref{F:xbgfeffort} measures the extraction effort and the involved grammar
transformations. Matching phases (\S\ref{X:nominal-matching} and
\S\ref{X:matching}) consist of semantics-preserving transformations, the
measurement for other phases is presented in the table directly. This
information was obtained in an automated manner but it relies on some amount of
semantic annotation of the transformations for the classifications and
convergence phases.

The number of transformations directly refers to the number of
\emph{applications of transformation operators}. As one can infer from
\autoref{F:xbgfpercommand}, 33 different operators are used in the JLS case;
most of them were introduced in \S\ref{S:transformation}. About three quarters
of the transformations are semantics-preserving. The remaining quarter is
mainly dedicated to semantics-increasing or -decreasing transformations with
only 2\% left for semantics-revising transformations.

In \autoref{F:xbgfeffort}, one can observe that relaxation
transformations indeed occur when a more readable and a more
implementable grammar are converged. Further, one can observe that the
overall transformation effort is particularly high for
\emph{jls12}---which signifies the fact (already mentioned above)
that \emph{impl1} and \emph{impl2} appear to be different
developments. Finally, we have made an effort to incorporate Oracle's bug
list into the picture (see ``Known bugs''). We note that some of the
known bugs equally occur in both the more readable and the more
implementable grammar, in which case we cannot even discover them by
grammar convergence.

\section{Related work}
\label{S:related}

We organize the related work discussion in the following manner:

\begin{itemize}
	\item grammar recovery (including grammar inference);
	\item programmable grammar transformations;
	\item other grammar engineering work;
	\item coupled transformations of grammar- or schema- or metamodel-like
artifacts and grammar- or schema- or metamodel-dependent artifacts;
	\item comparison (including matching) of schemas or metamodels.
\end{itemize}


\emphsec{Grammar recovery}

The main objective of the present JLS study is to discover grammar
relationships, but an ``important byproduct'' of the study is a consolidated
Java grammar\footnote{See also Software Language Processing Suite Grammar Zoo
at \url{http://slps.sf.net/zoo}.}. Hence, this particular instance of grammar
convergence (perhaps more than grammar convergence in general) relates strongly
to other efforts on grammar recovery. This topic has seen substantial interest
over the last decade because of the need for grammars in various software
engineering scenarios. We categorize this work in the following.


\subsubsection*{Recovery option 1: Parser-based testing and improvement cycle}

A by now classical approach to grammar recovery is to start from some sort of
documentation that contains a raw grammar, which can be extracted, and then to
improve the raw grammar through parser-based testing until all sources of
interest can be parsed (such as test programs, or entire software projects):
\citet{SellinkV00,LaemmelV01a,LaemmelV01b,JongeM01,AlvesV09}. The actual
improvement steps may be carried out manually
\citep{SellinkV00,JongeM01,AlvesV09} or by means of programmable grammar
transformations \citep{LaemmelV01a,LaemmelV01b}, as discussed in more detail in
\S\ref{R:trafo}.

The present JLS study, in particular, and the basic paradigm of
grammar convergence, in general, do not involve parser-based
testing. Instead, the similarity between two or more given grammars is
used as the criterion for possibly improving correctness.  Of course,
it would be a viable scenario to actually try deriving a useful parser
description from the converged Java grammar, and if additional
problems were found, then the parser-based testing and improvement
cycle of grammar recovery may be applied.


\subsubsection*{Recovery option 2: Grammar recovery from ASTs}

Generally, raw grammars (as discussed above) may also be extracted
from compilers. This is relatively straightforward, if the compiler
uses a parser description to implement the
parser. \citet{DuffyM07,KraftDM09} present another option, which
relies on access to the parse trees or ASTs of a compiler. A grammar
can be extracted from the ASTs for given sample programs. This
approach is specifically meant to help with the recovery of language
dialects for which precise grammars are often missing. In order to
derive the grammar for the concrete syntax, one must discover the
mapping between AST schema and concrete syntax. To this end, the
approach also involves some verification infrastructure. If we assume
that a baseline grammar is available (as opposed to a grammar for the
specific dialect at hand), then grammar convergence may also be useful
in providing the mapping between AST schema and concrete syntax.


\subsubsection*{Recovery option 3: Grammar inference}

Different authors have approached grammar recovery for software
languages through grammar inference techniques:
\citet{MernikGZB03,CrepinsekMJBS05,DubeyAJ05,DiPentaT05,DubeyJAS06,DubeyJA06,DiPentaLTT08,DubeyJA08}. Inference
relies on language samples, typically on both positive and negative
examples. Different inference scenarios have been addressed.
\citet{MernikGZB03,CrepinsekMJBS05} infer more or less complete
grammars, which is a very difficult problem. The approach applies to
small languages, e.g., small domain-specific languages.
\citet{DiPentaT05,DiPentaLTT08} start from a baseline grammar, and
infer modifications to the grammar so that all sources of interest can
be parsed. This search-based inference approach addresses the dialect
problem in software engineering, where a grammar for the language of
interest may be available, but not for the specific dialect at
hand. Both of the approaches use \emph{genetic
  algorithms}. \citet{DubeyAJ05,DubeyJAS06,DubeyJA06,DubeyJA08} use
a mix of advanced parsing and inference techniques instead.

Just as in the case of Option 1, the approach uses parser-based
testing as the correctness criterion, whereas grammar convergence
leverages the similarity between two or more given grammars as the
criterion for possibly improving correctness. It is quite conceivable
and interesting to combine grammar inference and grammar
convergence. For instance, grammar inference techniques could be used
to inform a semi-automatic grammar transformation approach. Also, it
is interesting to understand whether transformation operators for
convergence can usefully represent the modifications of the inference
approach of \citet{DiPentaT05,DiPentaLTT08}.


\subsubsection*{Recovery option 4: Special-purpose grammars}

Rather than trying to recover the (full) grammar for a given language,
one may also limit the recovery effort to specific samples, and more
potently, to the specific purpose of the grammar. For instance, when
the grammar is needed for a simple fact extractor, then there is no
need to parse the full language, or to be fully aware of the dialect
at hand. \citet{Moonen01,Moonen02} suggests so-called island grammars
to only define as much syntactical structure as needed for the purpose
and to liberally consume all other structure essentially as a token
stream. \citet{SynytskyyCD03} also pursue this approach specifically
in the context of multilingual parsing.  \citet{NierstraszKGLB07} also
pursue a variant of special-purpose grammars, where sample programs
are essentially modeled, and a grammar is computed from the
samples. A disciplined and productivity-tuned, iterative approach is
used to rapidly parse all the samples of interest. The approach also
produces the right metamodel (object model) to represent parse trees
tailored to the specific purpose at hand.


\emphsec{Programmable grammar transformations}
\label{R:trafo}

Grammar convergence, and some forms of grammar recovery, but also some
other software engineering problems rely on grammar transformations.
In fact, we would like to limit the focus here to \emph{programmable}
grammar transformations. We are not interested in ``hidden''
transformations as they may be performed implicitly by some software
tools such as a parser generator which removes left recursion
automatically.

Cordy, Dean and collaborators have invented the notion of agile
parsing \citep{DeanCMS03,Cordy03,DeanS05} and the paradigm of
grammar programming \citep{DeanCMS02} in this context. Both
concepts rely on language embedding of a grammar formalism into a
programming language (TXL, in their case). Agile parsing basically
suggests the customization of a baseline grammar for a specific use
case (such as components for reverse engineering or
re-engineering). The simpler programmable grammar transformations,
which are sufficient for some scenarios, are \textbf{redefine} (to redefine a
nonterminal), and \textbf{define} with the ability to extend the previous
definition.

In \citet{DeanCMS02}, a range of additional grammar programming
techniques is discussed, where some of these techniques can be
naturally modeled as grammar transformations (or more generally, as
program transformations). These are the techniques: rule abstraction
(so that grammar rules may be parametrized), grammar specialization
(so that the semantics of specific uses cases can be incorporated into
the grammar), grammar categorization (so that the resulting parser can
effectively deal with context-free ambiguities), union grammars (so
that one can have multiple grammars in the same namespace, perhaps
even with a non-empty intersection), and markup (i.e., the use of
markup syntax in combination with regular textual syntax).

In our own work (the one reported here, as well as in
\citet{Laemmel01a,LaemmelW01,GRK}), we have been interested in operator suites
for (programmable) grammar transformations. The idea is basically to view the
possible evolution of a grammar (along recovery or convergence) as a
disciplined editing process such that each editing step is described in terms
of an appropriate transformation operator. The use of an operator immediately
documents a certain intention, and is subject to precondition checking---just
like in other domains of program transformation. \citet{Wile97b} has also
suggested a small set of operators to specifically address the problem of
computing abstract from concrete syntax.


\begin{table}[p]
\begin{center}
\includegraphics[width=0.85\textwidth]{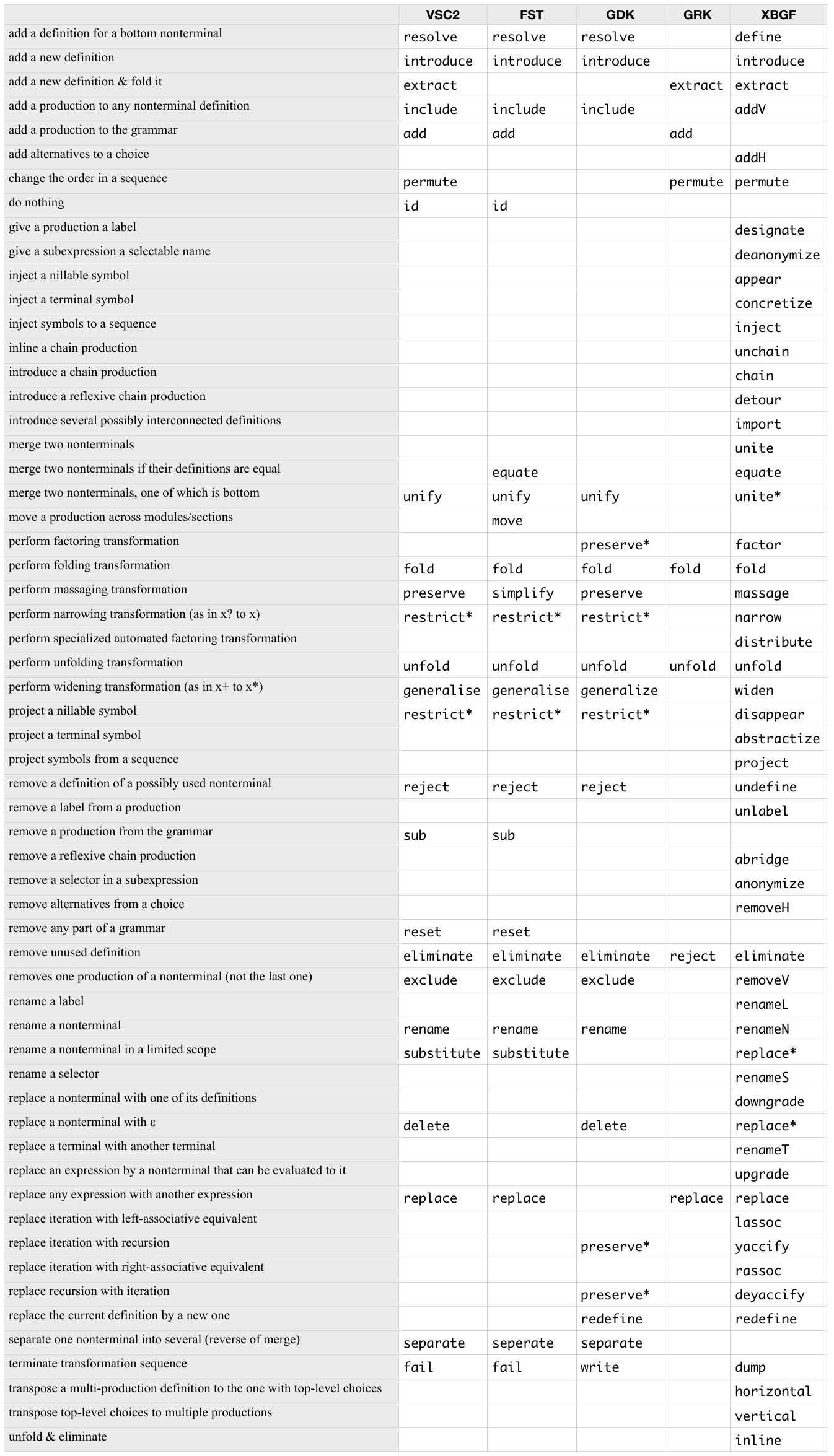}
\end{center}
\caption{Systematic comparison of grammar transformation operators provided by different frameworks}
\label{F:xbgf-vs-past}
\end{table}


\subsubsection*{The Amsterdam/Koblenz school of grammar transformation}

To better understand the design space of programmable grammar
transformations based on operator suites, we would like to compare
several efforts in which at least one of the authors has been
involved; see \autoref{F:xbgf-vs-past} for an overview. The figure
summarizes known grammar transformation operators, and compares
operator suites for grammar transformations:
\begin{description}
\item[VSC2~\citep{Laemmel01a,LaemmelV01b}] \mbox{}\\
The suite used for recovery of a Cobol grammar.\footnote{\url{http://homepages.cwi.nl/~ralf/fme01}}
\item[FST~\citep{LaemmelW01}] \mbox{}\\
A design experiment to define a comprehensive suite for
SDF \citep{VisserPhD}.\footnote{\url{http://www.cs.vu.nl/grammarware/fst}}
\item[GDK~\citep{GDK}] \mbox{}\\
A suite that is part of a grammar-deployment infrastructure.\footnote{\url{http://gdk.sf.net}}
\item[GRK~\citep{GRK}] \mbox{}\\
A suite that is part of an effort to reproduce
  our Cobol recovery case.\footnote{\url{http://slps.sf.net/grk}}
\item[XBGF] \mbox{}\\
The suite of the present paper; see \S\ref{S:xbgf}.\footnote{\url{http://slps.sf.net/xbgf}}
\end{description}
A starred name in the figure (as in ``restrict*'') means that the given
operator covers the function at hand, but it is more general.

XBGF, the transformation language of the present paper, provides clearly the
most comprehensive suite. There are a few empty cells in the XBGF column.
Reasons for non-inclusion differ; either the operator is considered too
low-level for the XBGF surface syntax (e.g., \textbf{substitute},
\textbf{reset}), or it is too low-level in the sense that all major application
scenarios are covered by more specialized operators (e.g., \textbf{add},
\textbf{sub}), or it is not currently implementable (e.g.,
\textbf{move}---modules are not fully supported in our infrastructure), or it
was simply not needed and perhaps debated so far (e.g., \textbf{delete},
\textbf{id}, \textbf{separate}, also known as FST's \textbf{seperate}; see the
table for the typo).

There is generally a tension between the number of transformation operators
vs.\ the achievable precision of a transformational program in terms of
expressing intentions, and thereby enabling extra sanity checks by the
transformation engine. Consider, for example, the line ``add a production to
the grammar''. This low-level idiom may be used to \textbf{include} another
production into an existing definition, or to add one or more productions in an
effort to \textbf{resolve} a missing definition, or to \textbf{introduce} a
definition for a so-far fresh nonterminal. In GRK, all these idioms are
modeled by \textbf{add}, and hence no intentions are documented, and no extra
checks can be performed automatically. In the case of XBGF, we have indeed
tried to separate idioms aggressively. This approach also helps us with
predicting the formal properties of each application of transformation
operators (i.e., semantics-preserving, -increasing, -decreasing, -revising),
and chains thereof.


\emphsec{Grammar engineering}

Let us also discuss some additional related work on grammar engineering
\citep{KlintLV05} in a broader sense. We begin with metrics which are used by
various recovery approaches and other work on grammar engineering. We want to
highlight \citet{AlvesV09,MalloyPW02,DuffyM07,JulienCFKMR09,KraftDM09}. Our
work leverages simple grammar metrics (numbers of bottom and top nonterminals)
and grammar-comparison metrics (numbers of nominal and structural differences)
for providing guidance in a grammar convergence context.

An interesting blend of recovery and convergence (or consistency checking) is
described in~\citep{BouwersBV08} where \emph{precedence rules} are recovered
from \emph{multiple} grammars and checked for consistency. At this point,
grammar convergence (in our sense) does not cover such sophisticated
convergence issues. In fact, our approach is, as yet, oblivious to
technology-specific representations of priority rules (as used in, say YACC or
SDF). We could potentially detect priority layers in plain grammars, though.

An alternative to grammar recovery is the use of a flexible parsing scheme
based on advanced error handling
\citep{Barnard81,BarnardH82,KlusenerL03}, subject to a baseline grammar.
Because of flexible parsing, the grammar could also be used to parse a dialect;
no precise grammar is needed. Also, code with syntax errors can be handled,
which is important in some application areas such as reverse or re-engineering
of legacy code.

There are approaches to connect the technical spaces of grammarware
and modelware in a manner that can be viewed as a form of grammar
convergence. That is, the parser may be obtained from the (meta)model
based on appropriate metadata and mapping rules, using a generative
approach \citep{JouaultBK06,NierstraszKGLB07}. We also use the term
model-driven parser development for these approaches. The point of
grammar convergence is that it provides a very flexible means to
represent relationships between grammar-like artifacts from different
technical spaces---without enforcing a particular scheme of designing
grammar-based artifacts or mappings.


\emphsec{Schema/metamodel comparison}

Grammar comparison, as it is part of grammar convergence, can be
loosely compared with \emph{schema matching} in ER/relational modeling
\citep{DoR07,RahmB01} as well as model and \emph{metamodel matching or
  comparison} in model-driven engineering
\citep{FalleriHLN08,WenzelK08,XingS06} (specifically in the context of
model/metamodel evolution). However, our current approach to
comparison (as of \S\ref{S:comp-trafo}) is relatively trivial, and
does not make any contribution to this subject, not even remotely. A
simple comparison approach was sufficient so far for two
reasons. First, the metamodel of grammars is relatively
simple. Second, we only require to determine nominal differences
(subject to the comparison of defined nonterminal names) and
structural differences (subject to matching alternatives). We will
need a more advanced comparison machinery once we aim at the partial
inference of grammar transformations. In this case, grammar
convergence should benefit from previous work on schema matching and
metamodel comparison.


\emphsec{Coupled transformations}

Grammar convergence relates to mappings in data processing
\citep{Thomas03b,LaemmelM06:XOR}, specifically to the underlying theory
of data refinement, and applications thereof 
\citep{Hoare72,Morgan90,ASVO05,Oliveira08,CunhaSV09}. In data
refinement, one also considers certain well-defined operators for
transforming data models. These operators must be defined immediately
in a way that they can be also interpreted as mappings at the data
level so that instance data can be converted back and forth between
the data models that are related by the transformation. 

Inspired by data refinement, all semantics-preserving and -decreasing
operators for grammar transformation can also be interpreted at the
AST level, and we experiment with such an interpretation, which opens
up new applications for grammar convergence. For instance, one could
replace the parser of a given program with another parser, even when
their AST types are different. That is, the convergence
transformations would be executed at the AST-level as a
conversion.

Data refinement is actually a specific and highly disciplined instance
of so-called coupled transformations, which are characterized to
involve multiple kinds of software artifacts (such as types vs.\
instance data vs.\ programs over those types) that depend on each
other in the sense that the transformation of one entity (of one kind)
necessitates a transformation of another entity (of another kind,
potentially) \citep{Laemmel04}. For instance,
\citet{HainautTJC93,LL01:RETIS,Wachsmuth07,VermolenV08,CicchettiREP08,BerdaguerCPV07}
are concerned with coupling for data models or metamodels vs.\
instance data or models; \citet{CleveH06} are concerned with coupling
for data models and programs over these data models. Again, we suggest
that grammar convergence should be generalized to cover coupled
transformations. As a result, the convergence method will find new
application areas.

\section{Concluding remarks}
\label{S:concl}

We have provided the first published record of recovering and
representing the relationships between given grammars of industrial
size that serve different audiences (language users and implementers)
and that capture different versions of the language. Our results
indicate that consistency among the different grammars and
versions---even for a language as complex as Java---is achievable.

The recovery and representation of grammar relationships is based on a
systematic and mechanized process that leverages a priori known
grammar bugs, grammar metrics (e.g., for problem indication), grammar
comparison for nominal and structural differences, and most notably,
grammar transformations. We carefully distinguish transformations for
grammar refactoring, extension, correction and relaxation.

While the JLS situation required the recovery of grammar
relationships, the ultimate best practice for grammar convergence
should require continuous maintenance of relationships. That is, the
relationships should be continuously checked and updated whenever
necessary along dependent or independent evolution of the involved
artifacts.

The approach, as it stands, faces a \emph{productivity problem}. The
transformation part of grammar convergence requires substantial effort
by the grammar engineer to actually map any given grammar difference
into a (short) sequence of applications of operators for grammar
transformation. For instance, the JLS transformations required several
weeks of just coding and debugging work. Such costs may be prohibitive
for widespread adoption of grammar convergence.

Notable productivity gains can be expected from advanced tool support.
We currently rely on basic batch execution of the transformations.
Instead, the transformations could be done interactively and
incrementally with good integration for grammar comparison,
transformation and error diagnosis.
Other productivity gains are known to be achievable by means of
normalization schemes (e.g., de-/yaccification in
\citet{JongeM01,Laemmel01a}).

However, ultimately, we need to provide inference of relationships (in
fact, transformations). Such inference is a challenging problem
because the convergence process involves elements of choice that we
need to better understand before we can promise reasonable
results. For instance, when two syntactic categories are equivalent
under fold/unfold modulations, then the grammar engineer is likely to
favor one of the two forms---this calls for either an interactive
approach or appropriate notions of normal forms or rule-based
normalization (i.e., heuristics).

Perhaps the most exciting remaining problem is to provide a proper formal
argument for the ``minimality'' of the non-semantics-preserving
transformations that are involved in a convergence. Currently, we use the
pragmatic approach to first align nonterminals, then to align alternatives
(by structure) as much as possible, and finally to break out of
refactoring and allow ourselves presumably local non-semantics-preserving
transformations. However, there is no formal guarantee currently for not
facing a false positive (``a presumed language difference that is none'').
That is, one may accidentally engage in semantics-revising transformations
even though the relevant syntactic categories are equivalent, but
nonterminal symbols or alternatives are confused by the grammar engineer.
Formally, the desired notion of minimality is limited by the
undecidability of grammar equivalence, but we are confident that a
practical strategy can be devised based on appropriate static analyses of
the transformations and the involved grammars.

Finally, a more strategic goal shall be to reconnect to
standardization bodies, and to examine potential for industrial
deployment of the method of grammar convergence. An early attempt,
preceding the development of grammar convergence, is documented in
\citet{NeedsGrammarware}. The challenge is here to understand what
sort of tools and refined methods would be acceptable for those users
in practice. This is an entirely non-trivial problem, but its solution
is critical to the value proposition of grammar convergence. Oracle, ISO,
and other stakeholders will not adopt grammar convergence tools and
methodology, unless they can measure the added value in terms of
productivity and correctness, and they do not need to engage with
uncomfortable dependencies on tool providers. For instance, they may
not like to completely overhaul their current methodology; they will
not use an experimental, academic, open-source project; neither will
they invest into major development of grammar-convergence tools; nor
will they hire a designated computer scientist with a PhD on grammar
engineering.


\begin{acknowledgements}
	The first author is grateful for
	opportunities to present and discuss some of this work on several
	occasions: University of Waterloo, IEEE Kitchener-Waterloo, August 6,
	2008; Dagstuhl Seminar 08331, ``Perspectives Workshop: Model
	Engineering of Complex Systems (MECS)'', August 12, 2008; \'Ecole des
	Mines de Nantes, November 7, 2009; METRIK Workshop, in Berlin,
	November 21, 2008; the BENEVOL Workshop (invited talk), Eindhoven,
	December 12, 2008; BX-Grace Meeting near Tokyo, December 15, 2008;
	DAIMI, University of Aarhus; February 24.
\end{acknowledgements}


\bibliographystyle{spbasic}      
\bibliography{paper}   

\newpage\appendix

\section{Grammar normalization}
\label{A:normalize}
If $(x,y)$ represents sequential composition of symbols $x$ and $y$, and
$(x;y)$ represents a choice with $x$ and $y$ as alternatives, then the
following formul\ae\ are used for normalizing grammars within our framework:

\begin{align*}
	(,)							& \Rightarrow \varepsilon				&	(;)					& \Rightarrow fail		\\
	(\ldots,(x,\ldots,z),\ldots)& \Rightarrow (\ldots,x,\ldots,z,\ldots)&	(x,)				& \Rightarrow x			\\
	(\ldots,x,\varepsilon,z,\ldots)& \Rightarrow (\ldots,x,z,\ldots)	&	(x;)				& \Rightarrow x			\\
	(\ldots;(x;\ldots;z);\ldots)& \Rightarrow (\ldots;x;\ldots;z;\ldots)&	\varepsilon^+		& \Rightarrow \varepsilon\\
	(\ldots;x;fail;z;\ldots)& \Rightarrow (\ldots;x;z;\ldots)			&	\varepsilon^\star	& \Rightarrow \varepsilon\\
	(\ldots;x;\ldots;x;z;\ldots)& \Rightarrow (\ldots;x;\ldots;z;\ldots)&	\varepsilon?		& \Rightarrow \varepsilon
\end{align*}

\section{Massage-equality}
\label{A:massage}
The massage-equality relation is defined by these algebraic laws:

\begin{align*}
	x? 		& = (x;\varepsilon)			& \qquad (x?)? 		& = x? 		& \qquad (x, x^\star) 	& = x^+		\\
	x? 		& = (x?;\varepsilon)		& (x?)^+ 			& = x^\star & (x^\star, x)	 		& = x^+		\\
	x^\star & = (x^+;\varepsilon)		& (x?)^\star 		& = x^\star & (x?, x^\star)	 		& = x^\star	\\
	x^\star & = (x^\star;\varepsilon)	& (x^+)? 			& = x^\star & (x^\star, x?) 		& = x^\star	\\
	x? 		& = (x?;x)					& (x^+)^+ 			& = x^+ 	& (x^+, x^\star) 		& = x^+		\\
	x^+ 	& = (x^+;x)					& (x^+)^\star 		& = x^\star & (x^\star, x^+) 		& = x^+		\\
	x^\star & = (x^\star;x) 			& (x^\star)?		& = x^\star & (x^+, x?) 			& = x^+		\\
	x^\star & = (x?;x^+) 				& (x^\star)^+ 		& = x^\star & (x?, x^+) 			& = x^+		\\
	x^\star & = (x?;x^\star) 			& (x^\star)^\star 	& = x^\star & (x^\star, x^\star) 	& = x^\star	\\
	x^\star & = (x^+;x^\star)
\end{align*}
$$x = (s_1::x; s_2::x)$$

\medskip

The infix operator ``::'' in the last formula denotes selectors (named
addressable subexpressions). They are needed because a choice between two
unnamed $x$ will always be normalized as $x$, as explained in
\autoref{A:normalize}.

\end{document}